\documentclass{article}

\usepackage{arxiv}

\usepackage[utf8]{inputenc} 
\usepackage[T1]{fontenc}    
\usepackage{url}            
\usepackage{booktabs}       
\usepackage{amsfonts}       
\usepackage{nicefrac}       
\usepackage{microtype}      
\usepackage{lipsum}		

\usepackage[utf8]{inputenc}
\usepackage[T1]{fontenc}
\usepackage{bm}
\usepackage{amsfonts}
\usepackage{amsmath}
\usepackage{amssymb}
\usepackage[numbers, sort&compress]{natbib}

\def\omg{{\Omega}}

\newcommand{\mcQ}{\mathcal{Q}}

\newcommand{\mcK}{\mathcal{K}}

\newcommand{\mcR}{\mathcal{R}}

\newcommand{\mcD}{\mathcal{D}}

\newcommand{\mcI}{\mathcal{I}}

\newcommand{\mcW}{\mathcal{W}}

\newcommand{\mcZ}{\mathcal{Z}}

\newcommand{\real}{\mathbb{R}}

\usepackage{graphicx}
\graphicspath{{fig/}} 
\usepackage{longtable}
\usepackage{tabularx}
\usepackage{arydshln}
\usepackage{geometry}
\usepackage{float}
\usepackage{placeins}
\usepackage[breakable]{tcolorbox}
\usepackage{enumitem}
\usepackage{xcolor}

\usepackage[colorlinks=true, citecolor=blue, linkcolor=blue]{hyperref}

\usepackage{fancyhdr}
\usepackage[ruled,vlined]{algorithm2e}
\usepackage{textgreek}
\usepackage{upgreek}
\usepackage{titlesec}
\titleformat{\paragraph}
{\normalfont\normalsize\bfseries}{\theparagraph}{1em}{}
\titlespacing*{\paragraph}
{0pt}{3.25ex plus 1ex minus .2ex}{1.5ex plus .2ex}
\titleformat{\paragraph}[runin]
{\normalfont\normalsize\bfseries}{}{0em}{}
\usepackage{color}
\definecolor{pink}{rgb}{1.0, 0.41, 0.71}

\definecolor{mygray}{rgb}{0.5, 0.5, 0.5}

\newcommand{\B}[1]{{\color[rgb]{0,0,1}#1}}

\usepackage{gensymb}

\usepackage{xr}

\title{Deep Neural Operator Enabled Concurrent Multitask Design for Multifunctional Metamaterials under Heterogeneous Fields}

\author{
  Doksoo Lee\\
  Department of Mechanical Engineering\\
  Northwestern University\\
  Evanston, IL 60208 \\
  \texttt{dslee@u.northwestern.edu} \\
   \And
 Lu Zhang \\
  Department of Mathematics\\
  Lehigh University\\
  Bethlehem, PA 18015\\
  \texttt{luz319@lehigh.edu} \\
  \AND
 Yue Yu \\
  Department of Mathematics\\
  Lehigh University\\
  Bethlehem, PA 18015\\
  \texttt{yuy214@lehigh.edu} \\
  \And
  Wei Chen\thanks{The corresponding author.} \\
  Department of Mechanical Engineering\\
  Northwestern University\\
  Evanston, IL 60208 \\
  \texttt{ weichen@northwestern.edu}   
}

\begin{document}


\maketitle











\begin{abstract}
Multifunctional metamaterials (MMM) bear promise as next-generation material platforms supporting miniaturization and customization. Despite many proof-of-concept demonstrations and the proliferation of deep learning assisted design, grand challenges of inverse design for MMM, especially those involving heterogeneous fields possibly subject to either mutual meta-atom coupling or long-range interactions, remain largely under-explored. To this end, we present a data-driven design framework, which streamlines the inverse design of MMMs involving heterogeneous fields. A core enabler is implicit Fourier neural operator (IFNO), which predicts heterogeneous fields distributed across a metamaterial array, thus in general at odds with homogenization assumptions, in a parameter-/sample-efficient fashion. Additionally, we propose a standard formulation of inverse problem covering a broad class of MMMs, and gradient-based multitask concurrent optimization identifying a set of Pareto-optimal architecture-stimulus (A-S) pairs. Fourier multiclass blending is proposed to synthesize inter-class meta-atoms anchored on a set of geometric motifs, while enjoying training-free dimension reduction and built-it reconstruction. Interlocking the three pillars, the framework is validated for light-by-light programmable plasmonic nanoantenna, whose design involves vast space jointly spanned by quasi-freeform supercells, maneuverable incident phase distributions, and conflicting figure-of-merits involving on-demand localization patterns. Accommodating all the challenges without a-priori simplifications, our framework could propel future advancements of MMM.

\end{abstract}


\section{Introduction}
Metamaterials are engineered architectured systems that support either superior often exotic functionalities, not found or beyond those in conventional systems~\cite{pendry2000negative}. The emergent behaviors usually stem from structure rather than composition~\cite{yu2018mechanical}. Among many, multifunctional metamaterials (MMM) form a branch of next-generation emerging material systems that support miniaturization and customization~\cite{lincoln2019multifunctional, wu2019mechanical, yuan2021recent}. They hold promise for making large societal impact, including that in 
medical, defense, and energy. A plethora of demonstrations on MMM have been reported in the literature~\cite{deng2018multifunctional, wu2019symmetry, li2019inverse, al2019multifunctional, chen2020metasurface, ze2020magnetic, tao2020multifunctional,  shirmanesh2020electro, an2021multifunctional, liu2021multifunctional}.

On one end, the recent advancements in fabrication techniques have fueled the dissemination of multiscale systems often possessing sophisticated architectural details~\cite{askari2020additive, bi2021all, yoon2016challenges, levchenko2018hierarchical, lichade2021hierarchical}. In contrast, design automation tools, particularly for MMM, are still underway. The lack of solid, rigorous design support has been, arguably, a root cause of the current design practice on MMM that largely resorts to a-priori simplifications on designable entities, e.g., material, topology of unit cells, tailorable incident light. Pre-specifying any of them not only risks a substantial compromise of what MMM could exclusively offer, but also impedes to shed light on the opaque inter-relationships among material, architecture, and stimulus~\cite{dolar2023interpretable}. In addition, a prevalent trade-off across multiple conflicting functionalities of MMM has rarely received in-depth investigations that it deserves. So has rigorous decision making under the trade-off, arguably due to the absence of a standard formulation of the inverse problem dedicated to MMM. 
Addressing all the design challenges all at once remains an open problem, while it has certainly been a core quest for design of next-generation metamaterials to embrace challenges, more than conventional approaches do, without losing tractability.

Since the inception of metamaterials~\cite{pendry2000negative}, a core quest of metamaterials design has been to achieve the sought-after ideal of the multiscale architectures through systematic, rational decision making. In doing so, recent years have seen a new wave of design paradigm: data-driven design for metamaterials and multiscale systems. The concept broadly refers to a design practice that builds on the common thread: discovering patterns from a finite collection of observational data, rather than domain knowledge, and then harnesses the patterns to expedite the otherwise arduous multiscale design procedure~\cite{lee2023data}. Evidenced by a growing volume of reviews from diverse perspectives~\cite{So2020DeepNanophotonics,  Regenwetter2021DeepReview, Kumar2021WhatMechanics, jin2022intelligent, woldseth2022use, so2022revisiting, liu2023deep, zheng2023deep, lee2023data}, the new paradigm has drawn immense attention in a broad realm of engineering sciences, perhaps with the vision that it may possibly unlock the potential of metamaterials.

The workhorse of the rising paradigm has been data-driven surrogates that offer on-the-fly, accurate enough prediction on quantities of interest, which by extension accelerates the resource-intensive, repetitive computations of solving governing equations for inverse design. The surrogate usually involves the mapping from a parameterized meta-atom to the corresponding homogenized, or ``aggregated'', spectral responses, e.g., scattering parameters, that are computed from the raw physical field. Grounded on the universal approximation theorem~\cite{cybenko1989approximation,chen1995universal,lanthaler2023nonlocal}, there always exists a data-driven model to approximate the associated mapping between input and output with a desired accuracy, given a sufficient amount of training data, low dimensional output, and fine-tuning of hyperparameters~\cite{lee2023data}. This camp of design approaches, which we would dub ``the palette approach''~\cite{Zhu2017} throughout this article, have underpinned foundational achievements reported in the communities~\cite{schumacher2015microstructures, panetta2015elastic, malkiel2018plasmonic, ma2019, liu2018generative, so2019designing, liu2022growth}.

Behind the proliferation, relatively sparse attention has been given to its intrinsic drawbacks. First, it typically resorts to the ``shortcut'' modeling strategy, where the model has access only to homogenized spectral responses at the downstream, which are at most a proxy of high-dimensional fields available in the full-wave analysis, e.g., rigorous coupled wave analysis~\cite{moharam1995formulation}, finite-difference time-domain~\cite{taflove2005computational}, and finite element method~\cite{jin2015finite}. Skipping the raw physical fields, the shortcut-based modeling hosts a disconnect from the domain knowledge that has been progressively forged by the communities, such as spatiotemporal causal effects formally described by either differential equations (e.g., the Maxwell's equations) or analytic models. Without proper supervision, the model would be black-box and prone to just imitating a superficial, statistical relation that is specific to given training data, rather than truly absorbing the fundamental light-matter interactions that are universal. Second, another underpinning of the palette approach is the assumption of scale separation, where (i) the multiscale system of interest is either deterministic under periodicity or stochastic following the ergodic principle thus stochastically periodic~\cite{lee2023data}, and (ii) uncoupled operation among non-identical neighboring meta-atoms can be justified with an acceptable amount of deviations from the spectral responses under periodic boundary conditions. The widely-used two conditions, however, are at odds with many advanced metamaterials that are subject to (i) long-range interaction~\cite{kwon2018nonlocal, overvig2020multifunctional, capers2021designing}, (ii) mutual coupling among non-identical meta-atoms~\cite{an2022deep, noh2022reconfigurable, ma2023incorporating}, (iii) fabrication uncertainty (e.g., heterogeneous local material defects; freeform meta-atom uncertainty)~\cite{wang2019robust, liu2020big, chen2022gan}, (iv) engineered disorder~\cite{jang2018wavefront, xu2022emerging, yu2021engineered}, (v) heterogeneously modulated incident wave~\cite{kao2012, lee2021dynamic, kim2022concurrent}. For those cases, the conceptualization, corroborated by follow-up experimental demonstrations, was presented a long time ago; yet the design automation with any optimality guarantee still remains open to questions. Developing a suite of inverse design tools dedicated to those systems could pave an avenue to further explore the rich design freedom in deterministic/stochastic multiscale architectures, succeeding to the recent achievements that have been primarily driven by the palette approach.

We claim that releasing the potential of MMM would in part be enabled via embracing the vast design space as is, jointly formed by multiple field-type design entities, followed by rigorous inverse optimization directly involving the set of on-demand functionalities at the system level, and their commonplace yet overlooked trade-off as well. To do so, we propose a data-driven design framework dedicated to MMM, with particular focus on programmable metamaterials that embody the multifunctionality of interest through a functional switch under modulation of external stimuli. The proposed framework builds on three pillars. First, we construct a field-to-field surrogate through implicit Fourier neural operator (IFNO) \cite{you2022learning,you2022physics,liu2023ino} (Section~\ref{Modeling: Field-to-Field Modeling through Implicit Fourier Neural Operator}). Neural operators~\cite{li2020neural,li2020fourier,lu2019deeponet,gupta2021multiwavelet,you2022nonlocal,kovachki2021neural,you2022learning,zhang2023metano} learn a surrogate mapping between infinite dimensional functional spaces thus feature resolution independence and generalizability to different input instances, which is in contrast to the classical neural networks, e.g., U-net~\cite{ronneberger2015u}, that only handle a mapping between finite-dimensional discrete vectors with prefixed resolution.  In our context, the neural operator offers the joint mapping from a pair of architecture (A) field (e.g., meta-atom and its arrangement) and stimulus (S) field (e.g., modulated incident phase), as the input instance, to the corresponding output field that could be heterogeneous beyond periodic boundary conditions. Once trained, the neural operator modeling approach offers on-the-fly full-field prediction, thus enjoys physical transparency, and generality to unseen field-type A-S pairs, all of which are challenging for the palette approach to achieve. 
Second, we present a standard formulation of the inverse problem on a class of MMM, where a system of interest are open to multiple types of designable entities and multiple target functional states (Section~\ref{Optimization An Inverse Problem Formulation on Multitask Concurrent Optimization of Multifunctional Devices}). Tackling the problem enables us to identify a set of Pareto-optimal architecture-stimuli pairs with respect to a user-defined set of functionalities, and to address task importance thereof, no matter if it has been given prior to the optimization or after. 
Third, we conceive Fourier multiclass blending (FMB) inspired by the topological encoding method for meta-atoms~\cite{liu2020topological}, and class remixing~\cite{chan2022remixing} in order to create a large amount of quasi freeform meta-atoms in relation to canonical families (Section~\ref{Data Generation: Meta-Atom Synthesis through Fourier multiclass blending}). The proposed meta-atom generation scheme allows to accommodate domain knowledge, produce inter-class instances specified through a compact yet expressive meta-atom representation that enjoys training-free dimension reduction and built-in reconstruction. 

To show the efficacy of the framework, design of spatially addressable plasmonic metasurfaces~\cite{singh2020far, buijs2021programming, lee2021dynamic, kim2022concurrent, yoo2023switching} is revisited as a case study, where the design goal is to embody a ``metasurface chessboard'' maneuverable with spatial light modulators (Section~\ref{Light-by-Light Programmable Plasmonic Metasurfaces}). As opposed to the prior reports that simplified the problem setting to make it amenable, we tackle all the core challenges simultaneously, namely
\begin{itemize}
    \item drastically varying fields involving heterogeneously distributed plasmonic hotspots
    \item opaque long-range interaction across meta-atoms
    \item the vast input space jointly formed by both a quasi-freeform metasurface supercell and spatially modulated phase field  
    \item a trade-off across the multiple functionalities of interest.
\end{itemize}

Through the case study, we also present the key findings including: 
\begin{itemize}
    \item the spatially drastically varying fields can be accurately predicted through the proposed IFNO architecture, with the local energy hotspots robustly reproduced 
    \item the machine learning (ML) surrogate combined with gradient-based multitask concurrent optimization boosts to navigate the vast stimulus-architecture joint space in a tractable way 
    \item inverse design of multifunctional devices forms a decision making problem on task importance specificity, which can be addressed by a set of diverse Pareto-optimal architectures, and one-to-many A-S solutions. 
\end{itemize}
In addition to the results in our case study, we delineate some feasible extensions of the proposed framework upon trivial modifications, in terms of alternative input representations for the neural operator, model transparency thereof, geometrically aperiodic metasurface arrays, and other optimization methods (Section~\ref{discussion}).

\section{Method}
\begin{figure}[!htb]
\centering
\includegraphics[width=0.9\textwidth]{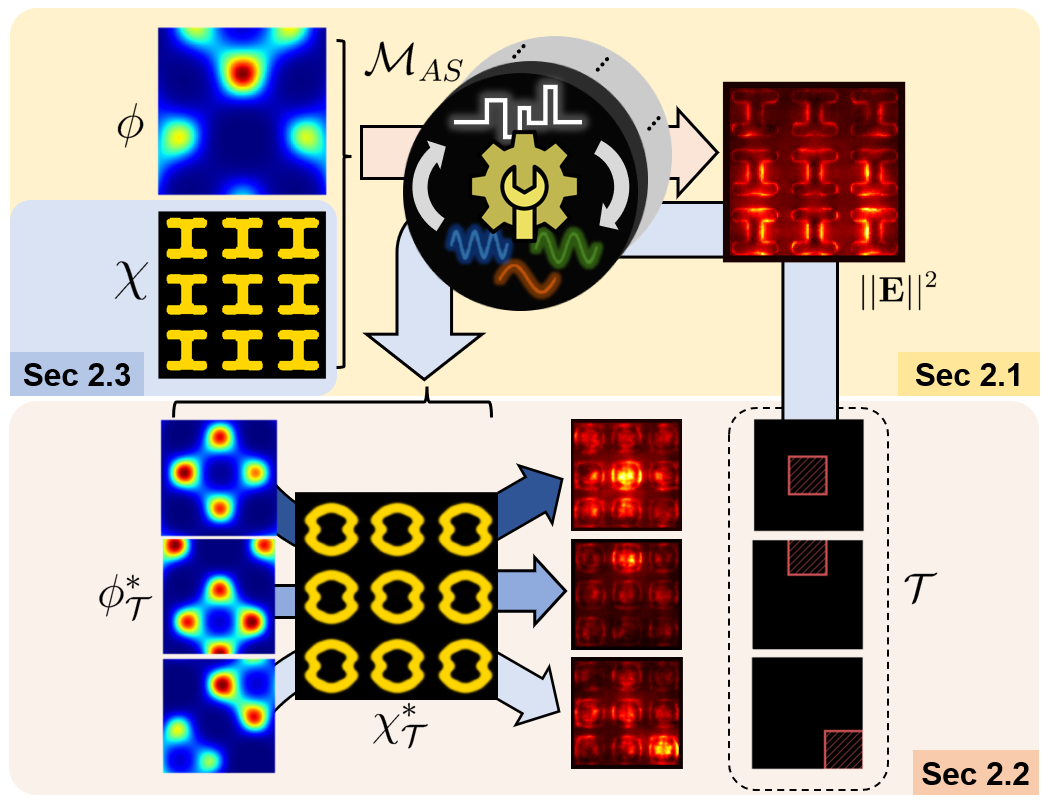}
\caption{A visual overview of the proposed framework.}
\label{overview_method}
\end{figure}

Figure~\ref{overview_method} gives a visual abstract of the proposed framework. Three key pillars are (1) fields-to-field modeling with implicit Fourier Neural Operator (Section~\ref{Modeling: Field-to-Field Modeling through Implicit Fourier Neural Operator}), (2) multitask concurrent A-S optimization (Section~\ref{Optimization An Inverse Problem Formulation on Multitask Concurrent Optimization of Multifunctional Devices}), and (3) quasi-freeform shape generation through Fourier multiclass blending (Section~\ref{Data Generation: Meta-Atom Synthesis through Fourier multiclass blending}).

    \subsection{Modeling: Field-to-Field Modeling through Implicit Fourier Neural Operator}
\label{Modeling: Field-to-Field Modeling through Implicit Fourier Neural Operator}
        \subsubsection{Operator Learning}
An operator refers to a mapping, or a function, that acts on elements forming a space and then generate elements forming another space, or optionally included in the same space~\cite{erdman2015functional}. The goal of operator learning is to learn a mapping between infinite-dimensional function spaces using a finite collection of observational data. As such, the trained operator can serve as an efficient surrogate for PDE solvers \cite{lu2019deeponet,li2020neural,li2020fourier}. Comparing with the classical PDE solvers and the physics-informed neural networks (PINNs), the neural operator approach requires no preliminary knowledge on the governing equation \cite{you2022physics} and, more importantly, is generalizable to different input instances. That means, once the network is trained, it provides on-the-fly predictions for new and unseen material architectures and stimulus.

We briefly state a formal description on operator learning. Let $\Omega \subset \mathbb{R}^d$ and $x \in \Omega$ be the spatial domain of the PDE and an arbitrary point within the domain, respectively. We consider two function spaces: (1) $A = A(\Omega; \mathbb{R}^{d_a})$ being a Banach space spanned by input functions taking values in $\mathbb{R}^{d_a}$ and (2) $U = U(\Omega; \mathbb{R}^{d_u})$ being another Banach space spanned by solution functions taking values in $\mathbb{R}^{d_u}$. The overarching goal of operator learning is to construct a (non-linear) mapping between the two function spaces by learning a solution operator $\mathcal{M}^{\dagger} : A \rightarrow U$, given a finite set of function pair observations $\{a_j(x), u_j(x)\}$, $x\in\Omega$. 
The ground-truth, infinite-dimensional map $\mathcal{M}^{\dagger}$ will be approximated through $\mathcal{M}$ endowed with parameterization $\theta \in \Theta$ following the mapping $\mathcal{M}_{\theta}: A \rightarrow U$, where $\Theta$ is the finite-dimensional parameter space for the operator. The optimal surrogate operator, $\mathcal{M}$, is then obtained by solving for  $\theta$ via the optimization problem: 
\begin{equation}
\theta^{\ast}=\underset{\theta \in \Theta}{\text{argmin}} \;\mathbb{E}_a[\mathcal{L}(\mathcal{M}_\theta(a), \mathcal{M}^{\dagger}(a))],
\end{equation}
where $\mathcal{L}(\cdot, \cdot)$ is the regression loss, e.g., mean squared error. When passing observational data to the neural operator, the collection would involve a finite discretization. Importantly, the learned operator $\mathcal{M}_{\theta^{\ast}}$ offers prediction with respect to \textit{any} $x \in \Omega$, including the points not necessarily covered by the finite discretization. This explicates why operator learning, as opposed to mainstream image-to-image networks like U-net~\cite{ronneberger2015u}, exclusively enjoys the discretization invariance: solutions can be transferred to different uniform grids (e.g., low-resolution $\rightarrow$ high-resolution). In addition, some reports claimed that operator learning also stands out regarding sample-/parameter-efficiency among its competitors~\cite{augenstein2023neural}.

        \subsubsection{Neural Operator}
Neural operator offers a powerful means to approximate non-linear operators with data. The general construction principle follows that of conventional, standard neural networks: taking a layer as a building block that involves some linear transformations (e.g., weight and bias), followed by nonlinear activations (e.g., ReLU or sigmoid), and then build a cascade of such blocks to boost modeling capability. While the standard neural networks capture the dependencies between neurons as discrete linear combinations, the integral neural operator (INO) architectures, as originally proposed by Li et al.~\cite{li2020neural} and extended in \cite{li2020fourier,you2022nonlocal,you2022learning,gupta2021multiwavelet}, model the long-range dependencies through an integral operator. An $L$-layer INO has the following form:
\begin{equation}\label{eqn:no}
\mathcal{M}_\theta[a;\theta](x):=\mathcal{Q}\circ\mcI_L\circ\cdots\circ \mcI_1\circ\mathcal{R}[a](x),
\end{equation}
where $\mcR$, $\mcQ$ are shallow-layer neural networks that provide point-wise mapping from a low-dimensional vector into a high-dimensional vector, and vice versa. Each intermediate layer, $\mcI_l$, consists of a local linear transformation operator $\mcW_l$, an integral (nonlocal) kernel operator $\mcK_l$, and an activation function $\sigma$. The architectures of NOs mainly differ in the design of their intermediate layer update rules. As a popular example, when considering the problem with structured domain $\omg$ and uniform grid set, the Fourier neural operator (FNO) is widely used \cite{li2020fourier}, where the integral kernel operators $\mcK_l$ are linear transformations in frequency space. The core idea of FNO is to parameterize the kernel integral operator $\mcK_l$ as a convolution operator defined in Fourier space. Formally, let $\mathcal{F}$ denote the FT of a function $f: \Omega \rightarrow \mathbb{R}^{d_v}$ and $\mathcal{F}^{-1}$ its inverse. FNO takes a sequence of Fourier layer blocks, such that the ($l$+1)th-layer feature function $v_{l+1}(x)$ is calculated based on the $l$th-layer feature function $v_l(x)$ via:
\begin{align}
&v_{l+1}(x)=\mcI^{FNO}_l[v_l(x)]:=\sigma(W_lv_l(x)+\mathcal{F}^{-1}[\phi_l\cdot \mathcal{F}[v_l(\cdot)]](x)+ c_l),\label{eq:FNO}
\end{align}
where $W_l$, $c_l$ and $\phi_l$ are trainable matrices to be optimized. Comparing to other NO architectures, FNO inherits the advantages of INOs on resolution independence while boosting the computation of the kernel integral operator $K$ using Fast Fourier Transform (FFT). The approximation capabilities have been demonstrated by a diverse array of benchmarks~\cite{wen2022u, you2022learning, lu2022multifidelity, augenstein2023neural}.


        \subsubsection{Implicit Fourier Neural Operator}
        \label{IFNO_description}

Given sufficient amount of data, FNOs are universal in the sense that they can approximate any continuous operator to a desired accuracy \cite{kovachki2021neural}. However, in the vanilla FNO architecture \eqref{eq:FNO} different integrable layers have different parameters $W_l$, $c_l$, and $\phi_l$, which makes the number of trainable parameters increasing as the network gets deeper. As a result, the training process of the FNOs may become challenging and potentially prone to over-fitting in small data regimes \cite{you2022nonlocal,you2022learning}. To overcome these limitations, the implicit neural operator architectures were proposed \cite{you2022learning,li2023modeling,benitez2023fine}, where the skip connections were added between the integral layers and all layers share the same set of trainable parameter, $W$, $c$, and $\phi$. In particular, the architecture 
Eq.~\ref{eq:FNO}
was modified as:
\begin{align}
&v_{l+1}(x)=\mcI^{IFNO}[v_l(x)]:=v_l(x)+\dfrac{1}{L}\sigma(Wv_l(x)+\mathcal{F}^{-1}[\phi\cdot \mathcal{F}[v_l(\cdot)]](x)+ c).\label{eq:IFNO}
\end{align}
As a result, the NO architecture has substantially smaller number of trainable parameters, while still serves as a universal approximator for fixed point PDE solvers \cite{you2022learning}. Moreover, the above architecture can also be interpreted as discretized nonlocal differential equations, and consequently allows for the shallow-to-deep initialization technique where optimal parameters learned on shallow networks are considered (quasi-optimal) initial guesses for deeper networks \cite{you2022nonlocal}. This technique was found helpful in enhancing the network consistency across different layer numbers, and mitigating the vanishing gradient issue in the deep layer limit \cite{haber2018learning,you2022nonlocal}.

In this paper we consider $\Omega$ a squared-shape 2D measuring domain with uniform grids, and employ the implicit Fourier neural operator (IFNO) \cite{you2022learning} as the surrogate solver. In practice, the implementation of FNO and its variants is accelerated through FFT that incurs quasilinear time complexity, provided that the discretization over $\Omega$ is uniform. Otherwise, use of special strategies dedicated to unconventional domains~\cite{li2022fourier,liu2023domain} was proposed; this beyond the scope of the current work.

Herein we introduce the architectural details of the IFNO network $\mathcal{M}_{AS}$ conceived for this work. The concatenated vector of 2D coordinate $x$, the architecture function, and the stimulus function is set as the input field, and the corresponding electric field as the output function. As a result, we have the input function $a$ being vector-valued with $a(x)\in \real^4$, and the output $u$ being scalar-valued functions, i.e., $u(x)\in\real$. In the proposed network architecture, key model parameters that primarily determine the model complexity are as follows:
\begin{itemize}
    \item $\text{modes}$: the number of Fourier modes along each dimension
    \item $\text{width}$: the number of channels
    \item $\text{lastwidth}$: the number of input channels at the last fully connected linear layer
    \item $\text{depth}$: the number of cascaded Fourier layers 
\end{itemize}
Through the parameter study on the model parameters, we use $\text{modes}$=64, $\text{width}$=32, $\text{lastwidth}$=64, and $\text{depth}$=8. Parameter study on depth can be found in Appendix \textcolor{blue}{A}. The setting forms 8,392,001 parameters, all of which are trainable. Hyperparameter setting includes $lr=0.01$ as initial learning rate, $\gamma=0.5$ as scheduler exponent, $wd=10^{-5}$ as weight decay in the Adam optimizer~\cite{kingma2014adam}, $n_{epochs}=500$ as the number of epochs, and $n_{step}=100$ as the period of step to update the scheduler. All trainings are performed on nVIDIA Tesla T4 GPU card with 16GB memory. Further details associated with the training, e.g., scheduling~\cite{goodfellow2016deep}, 
 are stated in Appendix \textcolor{blue}{A}.



    \subsection{Optimization: An Inverse Problem Formulation on Multitask Concurrent Optimization of Multifunctional Devices}
    \label{Optimization An Inverse Problem Formulation on Multitask Concurrent Optimization of Multifunctional Devices}    
        \subsubsection{The Inverse Problem on Programmable Material Systems}
The end goal of this work is to present an inverse design framework for multifunctional systems that: (1) exhibit an on-demand functional switch subject to dynamic stimuli and (2) are possibly open to multiple types of design entities, e.g., architecture (A), material (M), and stimulus (S), not limited to architecture (i.e., meta-atom and its tessellations). The proposed design framework is supposed to navigate the vast design space, jointly formed by M-A-S entities, with a possibly conflicting target tasks addressed along the way. More specifically, we are interested in A-S concurrent optimization for plasmonic metasurfaces that are programmable under spatial phase modulation~\cite{ singh2020far, buijs2021programming, lee2021dynamic, kim2022concurrent, yoo2023switching}, which offer a means to embody multifunctionality with a geometrically stationary metasurface. In addition, the presented framework is applicable for more than two target functional states, as opposed to a plethora of demonstrations done for on-off type programmability~\cite{buchnev2015electrically, li20214d, wang2023physics, wang2023inverse, malek2017strain, bilal2017reprogrammable, xu2019stretchable}, mostly presented without the trade-off investigated. The case study of main interest that involves all the challenges is to be elaborated in Section~\ref{Light-by-Light Programmable Plasmonic Metasurfaces}.

        \subsubsection{A Generic Formalism for Multitask Architecture-Stimulus Concurrent Optimization}
As one form to support multifunctionality, a functional switch in programmable material systems involves a finite set of target states, each of which can generally be conceptualized as \textit{task}. For a single task $t$, let $\theta_A \in \Theta_A$ be the parameterization specifying architecture, e.g., a set of bar lengths for I-beam meta-atoms, $\theta_S\in \Theta_S$ be that given to stimulus, and $\mathcal{J}_{t}: \Theta_A \times \Theta_S \rightarrow \mathbb{R}$ be the task-specific objective that quantifies system performance with regard to task $t\in \mathcal{T}=\{1, \cdots,  T \} \subset \mathbb{N}$. For each task, the A-S design problem can be formulated as:
\begin{equation}
\label{eq:single task loss}
    ((\theta_{A})_t^{*}, (\theta_{S})_t^{*})=\min_{ (\theta_{A}, \theta_{S}) \in \Theta_A \times \Theta_S }{\mathcal{J}_{t}(\theta_{A}, \theta_{S})},
\end{equation}
where the superscript $(\cdot)^{*}$ and subscript $(\cdot)_{t}$ indicate optimality and the associated task index, respectively. Given a single task $t$ in concern, an A-S pair $((\theta_{A})_t^{*}, (\theta_{S})_t^{*})$ can be optimized via navigating the joint $\Theta_A \times \Theta_S $ space. By extension, given a set of tasks, optimized A-S pairs can be found individually. Seemingly straightforward, this formulation Eq.~\ref{eq:single task loss} turns out to be ineffective for real-world deployment of multifunctional devices. It is highly likely that optimized A-S pairs for individual tasks would involve task-specific optimal architectures; demanding different architectures for different functionalities would hinder the practical deployment. It is warranted to find a single architecture that supports multiple functionalities under dynamic stimuli. To do so, we alternatively consider the following optimization problem:
\begin{equation}
\label{eq:multiple task loss}
    \bm{\uptheta}^* = (\theta_{A}^*, \bm{\uptheta}_{S}^*)=\min_{  \theta_{A} \in \Theta_A, \theta_{S} \in \Theta_S }\mathbf{J}(\theta_{A}, \bm{\uptheta}_{S}) = 
    \min_{   \theta_{A} \in \Theta_A, \theta_{S} \in \Theta_S }\mathbf{J}(\theta_{A}, (\theta_{S})_1, \cdots, (\theta_{S})_T)
\end{equation}
where $\bm{\uptheta}$ is the concatenated A-S entities; $\bm{\uptheta}_{S}=\{ (\theta_{S})_1, \cdots, (\theta_{S})_T \}$ denotes the set of stimuli; $\mathbf{J}$ is the multitask objective, formulated as a vector-valued function of individual figure-of-merits (FoMs) $\mathcal{J}_{t}$. The way to specify $\mathbf{J}$ with respect to multiple FoMs is non-singular. A conventional, empirical approach is to construct the so called \textit{scalarized} objective as a proxy, which casts the multiobjective optimization problem into a single-objective one as follows:
\begin{equation}
\label{eq:scalarized objective}
    \min_{ (\theta_{A}, \theta_{S}) \in \Theta_A \times \Theta_S } \sum_{t=1}^{T} {c_{t} \mathcal{J}_{t}( \theta_{A}, \bm{\uptheta}_{S} )},
\end{equation}
where $0\leq c_{t} \leq 1 $ is the weighting factor subject to the linear constraint $\sum_{t=1}^{T} c_{t} = 1$. Provided that all the FoMs have been properly scaled one another, the key is how to assign the scaling $c_{t}$ under the constraint. A large body of prior work has formed a dichotomy between static weighting where $c_{t}$ remains constant during optimization and dynamic weighting based on some heuristics~\cite{ marler2010weighted}. Whichever is chosen, it is inevitable for the weighted formulation to involve a grid search or heuristics unless task importance has been declared. Such is rarely the case, hence the line of approaches allows room for subjectivity in terms of global optimality.

The optimization goal of Eq.~\ref{eq:multiple task loss}, for generic purposes, should ideally be to find a set of solution(s) $\mathcal{P}_{\theta}$ that meets Pareto optimality~\cite{sener2018multi}. The concept builds on the notion of a \textit{dominated solution}: $\theta^{*}$ is said to dominate $\theta$ if $\mathcal{J}_t(\theta^{*}) \leq \mathcal{J}_t(\theta)$ for $ \forall t \in \mathcal{T}$ and $\mathcal{J}_t(\theta^{*}) \neq \mathcal{J}_t(\theta)$. No solution $\theta$ dominates a Pareto optimal solution $\theta^{*}$. Parato optimal solutions constitute the Pareto set $\mathcal{P}_{\theta}$, whose image with infinitely many elements would form the Pareto front $\mathcal{P}_{\mathbf{J}}$.

The Pytorch implementation of the proposed IFNO (Sec.~\ref{IFNO_description}) offers access to numerical design sensitivities of $\mathcal{M}_{AS}$ through backpropagation with Automatic Differentiation~\cite{paszke2017automatic} (Figure~\ref{backprop}). Capitalizing on the gradients, we tackle the optimization problem in Eq.~\ref{eq:multiple task loss} based on gradient descent. Analogous to the conventional optimization with a single objective, gradient descent for multiobjective optimization boils down to finding a Pareto stationary point, defined based upon the Karush-Kuhn-Tucker (KKT) conditions. Pareto stationary points can be found through the following line search~\cite{sener2018multi}:
\begin{equation}
\begin{aligned}
\min_{ \{ c_{t} \} } \left( \left\| \sum_{t=1}^{T} c_{t} \nabla_{\theta_{ A }} \mathcal{J}_t(\theta_{A }, \bm{\uptheta}_{S}) \right\|^2 \right), 
\end{aligned}
\label{eq: line search}
\end{equation}
under the aforementioned constraints on the weights $\{ c_{t} \}$. It has been shown that the minimizer of Eq.~\ref{eq: line search} either ensures that the gradient direction decreases all task-specific objectives $\mathcal{J}_t$ or meets the KKT conditions with the minimum norm being zero~\cite{sener2018multi}. Built on the availability of an analytic solution for $|\mathcal{T}|=2$, a common descent direction for $|\mathcal{T}| \geq 3$ can be identified. For each iteration, the design update with respect to stimulus, i.e.,  ($\theta_{S})_{t} \rightarrow (\theta_{S})_{t}'$, is task-specific hence done independently across tasks as follows:
\begin{equation}
    (\theta_{S})_{t}' = (\theta_{S})_t-\eta_{S}
    \nabla_{(\theta_{ S})_t} \mathcal{J}_t(\theta_{A }, \bm{\uptheta}_{S}),
\label{eq: design update for stimulus}
\end{equation}
with the step size on stimulus being $\eta_S$. On the other hand, design update for architecture is shared across all tasks. The design update for architecture ($\theta_{A} \rightarrow \theta_{A}'$) -- the most challenging part in this optimization -- is specified via aggregating task-specific gradients through
\begin{equation}
    \theta_{A}' = \theta_{A}-\eta_{A}
    \sum_{t=1}^{T} c_{t} \nabla_{\theta_{ A }} \mathcal{J}_t(\theta_{A }, \bm{\uptheta}_{S}),
\label{eq: design update for architecture}
\end{equation}
where $\eta_A$ is the step size for architecture, and $\{c_t\}$ is identified via the line search (Eq.~\ref{eq: line search}). The proposed formalism will be employed in the case study in Section~\ref{Light-by-Light Programmable Plasmonic Metasurfaces}, where modulated incident phase and meta-atom geometry are regarded as stimulus and architecture, respectively.

    \subsection{Data Generation: Meta-Atom Synthesis through Fourier multiclass blending}
\label{Data Generation: Meta-Atom Synthesis through Fourier multiclass blending}        
In data-driven design, data itself is a design element~\cite{lee2023t}. Data acquisition for D3 has been tackled through a diverse array of strategies~\cite{Regenwetter2021DeepReview, lee2023data, zheng2023deep}. Lee et al.~\cite{lee2023data} claimed that what commonly underlies each were (1) unit cell representation, e.g., pixel/voxel, and (2) reproduction strategy, e.g., parametric sweep. A representation refers to a set of parameters, or models, used to directly characterize unit cells~\cite{Chan2022Yu-ChinDissertation}. Meanwhile, reproduction refers to the way of producing generic shape instances, particularly of ``growing'' a sparse shape set to massive one~\cite{lee2023data}. Determining the pair of representation and reproduction, specifically for the meta-atoms within our scope, is a key decision that should be made at the early stages of data acquisition, as the pair (1) dictates the distributional nature of resulting data, e.g., space-filling and coverage, and (2) could significantly affect the difficulty of downstream tasks, e.g., both unit-cell-level and system-level design optimization. 

        \subsubsection{Fourier multiclass blending}
Harnessing the Fourier Transform 
pair, Liu et al.~\cite{liu2020topological} has established a procedure to obtain a versatile design representation that is powerful to specify topologically free meta-atoms. The key idea is that given an image of a meta-atom, one can apply $\mathcal{F}$ to obtain the corresponding sparse representation in the frequency domain. The method enjoys (1) substantial dimension reduction with topological skeletons preserved, (2) built-in training-free reconstruction capability supported by $\mathcal{F}^{-1}$, and (3) perfect control over some topological symmetry of meta-atoms.
       
Hinged on the benefits, we present FMB. The core motivation of multiclass blending~\cite{chan2022remixing}, in general, is to generate a large amount of building blocks, pivoted on a set of geometric motifs, as many as needed. We advocate this line of reproduction strategy for D3 purposes in that it directly accommodates domain knowledge via including canonical meta-atom families and smoothly bridges them, often in a unified landscape with follow-up representation learning~\cite{Chan2022Yu-ChinDissertation}. In doing so, we propose to do multiclass blending in the Fourier feature space in order to promote smooth interpolation between an arbitrary pair of meta-atom instances. This is a departure from Chan et al.~\cite{chan2022remixing} where the blending takes place in the ambient image space. Empirical observations support that the proposed blending offers a smoother transition (Figure~\ref{SI_FMB_smooth_comparison}, Appendix \textcolor{blue}{C}).

We assume that FMB starts from a predefined set of $nc$ families, whose individual ``class representative'' is given as either a binary image or a level-set function ${\chi}(x,y) \in [0, 1]^{N\times N}$. For each binary image, a corresponding sparse Fourier representation $\boldsymbol{z}_k (k=1, 2, \cdots, nc)$ can be found with FT $\mathcal{F}$, where $dim(\boldsymbol{z}_k) \ll N^2 $. The feature dimension is empirically determined by observing the trade-off between dimensionality $dim(\boldsymbol{z}_k)$ and reconstruction error (Figure~\ref{SI_FMB_error_historgram}, Appendix \textcolor{blue}{C}). Since the Fourier features involve highly imbalanced distributions across components, it is useful to apply proper feature-wise scaling, e.g., standardization or normalization. We create an inter-class instance through a simple linear combination of the features in the Fourier feature space as   
    \begin{equation}
         \boldsymbol{z} = \sum_{k=1}^{n_b} \alpha_k \boldsymbol{z}_k,
    \label{eq:MFB}
    \end{equation}
where $n_b \geq 2 \in \mathbb{N}$ is the number of classes to be considered for blending, as $\alpha_k$ the weights subject to $\sum_{k=1}^{n_b} \alpha_k =1$ and $0 \leq \alpha_k \leq 1$. For reconstruction, the corresponding level-set representation is found through:
    \begin{equation}
    \label{eq:FMB_1}
    \hat{\phi}(x,y) =  \mathcal{F}^{-1} [ \boldsymbol{z} ].
    \end{equation}
Given a cutoff threshold $\phi_0$, the binary image of the inter-class meta-atom is identified as 
    \begin{equation}
    \hat{\chi}(x,y) =     
    \begin{cases}
    1 & \text{where}\ \hat{\phi}(x,y) \geq \phi_0 \\
    0 & \text{otherwise}. \\
    \end{cases}    
    \label{eq:FMB_thresholding_from_phi}
    \end{equation}

\begin{figure}[!htb]
\centering
\includegraphics[width=0.9\textwidth]{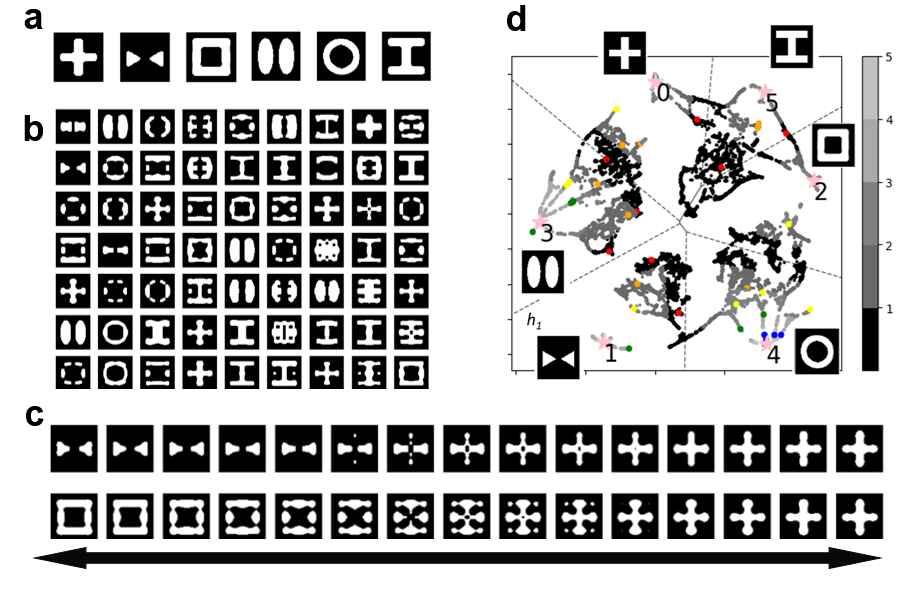}
\caption{
An overview of the proposed FMB and its instantiation $ \mcD_S$. (a) The six start-up classes chosen from literature. (b) Seventy randomly selected inter-class instances. (c) Two-class linear traversal in the 36D Fourier feature space $\mcZ$. (d) A 2D shape manifold obtained through Uniform Manifold Approximation and Projection~\cite{mcinnes2018umap}.
}
\label{FMB_shape_examples}
\end{figure}
   
Figure~\ref{FMB_shape_examples} shows example instances of the proposed blending. Without loss of generality of the blending scheme, six classes were selected from literature (Figure~\ref{FMB_shape_examples}(a)). By construction, all instances generated by FMB inherits the advantages of the sparse Fourier representation, i.e., dimensional compactness, built-in reconstruction, and symmetry control. The proposed blending offers smoother transition between/among user-defined class representatives, and inter-class instances born from those (Figure~\ref{FMB_shape_examples}(c)). Based on the FMB approach, we built a ground shape set, to be called $\mathcal{D}_{S}$ throughout this article, including 15k instances with $n_b=2, 3, 4$. In Figure~\ref{FMB_shape_examples}(b), some instances show clear closeness to one of the user-define families, e.g., bow-tie, ellipse, I-beam. Most of the instances, however, exhibit ``organic'' deviations from the families, which are difficult to be explicitly described through simple primitives, e.g., bar and hole. More instantiations of for two- and three-class blending are listed in Figure~\ref{SI_FMB_example_instances} of Appendix \textcolor{blue}{C}. Figure~\ref{FMB_shape_examples}(d) visualizes a 2D shape manifold of the unified landscape obtained through Uniform Manifold Approximation and Projection~\cite{mcinnes2018umap}. The visualization inevitably distorts the original 36D Fourier feature space; yet some qualitative observations are available, e.g., the closeness between bow-tie family and ellipse family, that between ring and square ring. The color denotes the distance to the nearest class representative, which could serve as a rough metric to quantify the degree of blending.

\section{Case study: Light-by-Light Programmable Plasmonic Metasurfaces}
\label{Light-by-Light Programmable Plasmonic Metasurfaces}
Active light control using plasmonic metasurfaces paves a way to enhance the speed of nanoscale optical imaging using far-field optical components~\cite{shaltout2019spatiotemporal, kang2019recent}. Spatial phase modulation~\cite{singh2020far, buijs2021programming, lee2021dynamic, kim2022concurrent, yoo2023switching} offers a route to dynamically address light confinement on plasmonic metasurfaces~\cite{balogun2019optically}. The inverse problem involves two key designable entities: plasmonic meta-atoms (Section~\ref{Data Generation: Meta-Atom Synthesis through Fourier multiclass blending}), whose array channels propagating light into localized evanescent electromagnetic waves on the sample surface, and phase distributions in the incoming wave (Section~\ref{phase representation}), whose distribution can be modulated through commercial spatial light modulators.

    \subsection{Phase Representation} 
\label{phase representation}    
Based on our perspective for inverse design, phase is a special field-type instance of stimulus that is open to design. 
As a proof-of-concept of the proposed framework, we employ the continuous phase representation presented in Lee et al.~\cite{lee2021dynamic}:
\begin{equation}
\begin{aligned}
\phi(x, y) & = \sum_{j=1}^{n_h} \phi_j \Phi_j (x, y) \\
\Phi_j (x, y) & = \phi_j \cos\left( \frac{\pi}{j\lambda} Mx + \alpha_j\right) \cos\left( \frac{\pi}{j\lambda} My + \beta_j\right)
\end{aligned}
\label{eq:harmonic_phase_representation}
\end{equation}
where $0 \leq \phi_j \leq \pi, \, 0 \leq \alpha_j, \beta_j \leq 2\pi$; $\phi$ is the spatial phase lag function; $n_h$ is the order of the expansion; $\phi_j$ represents the amplitude of harmonic $j$; $\alpha_j$ and $\beta_j$ establish a translational shift of harmonic $j$ along the $x$- and $y$-directions, respectively; $\lambda$ is the periodicity of a meta-atom; $M$ is the demagnification factor. An individual harmonic is specified by three design variables $\left[\phi_j, \alpha_j, \beta_j\right]^T$. With $n_h=2$ a phase profile is represented by six design variables $\bm{\upphi}=\left[\phi_1, \alpha_1, \beta_1, \phi_2, \alpha_2, \beta_2\right]^T$. Further details of the full-field analysis using COMSOL Multiphysics\textsuperscript{\textregistered} v.5.6
~\cite{comsol2020} can be found in either Lee et al.~\cite{lee2021dynamic} or Appendix \textcolor{blue}{B}. Note that the proposed framework involving IFNO (Section~\ref{Modeling: Field-to-Field Modeling through Implicit Fourier Neural Operator}) is by no means restricted to this particular analytic representation.




    \subsection{The Inverse Problem on Digitally Addressable Plasmonic Metasurfaces}   
\begin{figure}[!htb]
\centering
\includegraphics[width=0.8\textwidth]{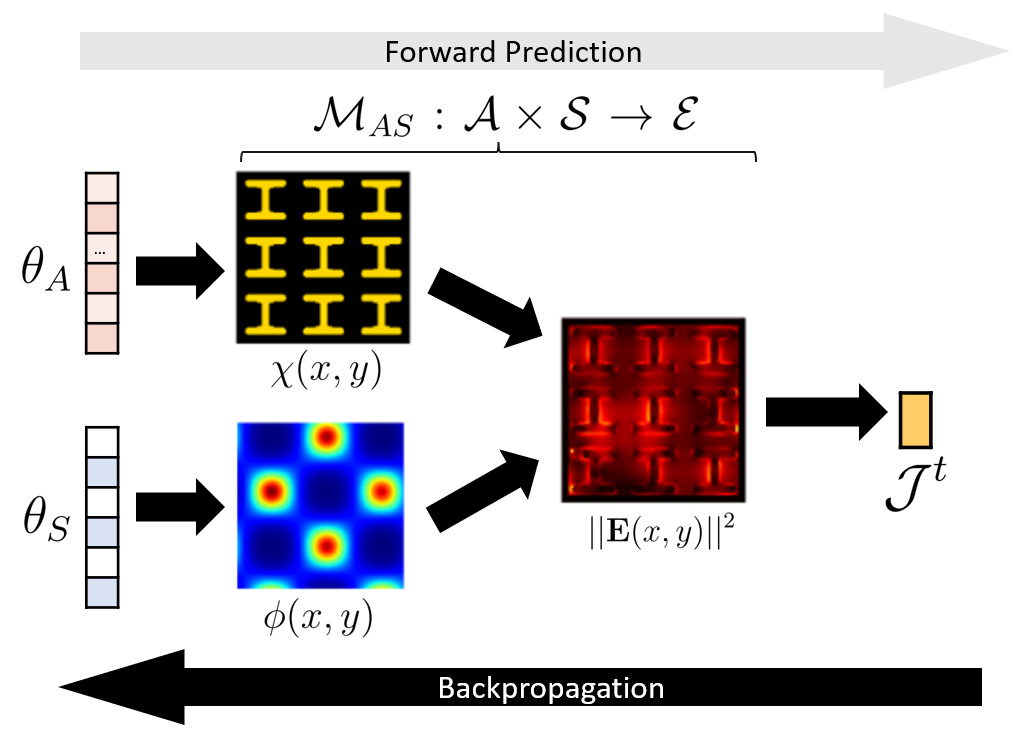}
\caption{An illustration of backpropagation in $\mathcal{M_{AS}}$ to obtain the numerical gradients.}
\label{backprop}
\end{figure}

The full-wave analysis offers an access to norm of electric field $||\mathbf{E}(x, y)||$ for $(x,y)\in \Omega$, where $\Omega$ is the measuring domain located right above the metasurface array. The planar measuring domain is subdivided into $N_m \times N_n$ square patches of equal size, each of which is denoted as $\Omega_{(m, n)}$ for $m, n=1, 2, \cdots, N_m$. We consider the following array response matrix $\mathbf{L}=[L]_{(m,n)}$ given a plasmonic metasurface antenna array:
\begin{equation}
L_{mn} = \int_{\Omega_{(m,n)}}||\mathbf{E}(x,y)||^2  d\Omega.
\end{equation}
As a proof-of-concept, we set the three patterns as the target task of multitask concurrent A-S optimization (i.e., $\mathcal{T}=\{1, 2, 3$\}). In order to construct a scalar FoM specific to each energy redistribution pattern, task-specific weight matrices $\mathbf{W}_{t}$ are introduced as:
\begin{equation}
    \begin{aligned}
    \mathbf{W}_{1}=\left[ -1, -1, -1; -1, 8, -1; -1, -1, -1 \right] \\
    \mathbf{W}_{2}=\left[ -1, 8, -1; -1, -1, -1; -1, -1, -1 \right] \\ 
    \mathbf{W}_{3}=\left[ -1, -1, -1; -1, -1, -1; -1, -1, -8\right].       
    \end{aligned}    
\end{equation}
An intuitive visualization will be provided in Section~\ref{Multitask Concurrent Optimization}. With $\mathbf{W}_{t}$ specific to a target localization pattern, an individual FoM $\mathcal{J}_{t}$ is quantified as
\begin{equation}
\mathcal{J}_{t} = \mathbf{W}_{t} \circ \mathbf{L}   
\end{equation}
where $\circ$ is the Hadamard product (i.e., elementwise multiplication). Deduced from Eq.~\ref{eq:multiple task loss}, the associated inverse problem reads
\begin{equation}
    \min_{ (\boldsymbol{z}, \bm{\upphi}) \in \Theta_A \times \Theta_S } \sum_{t=1}^{T} {c_{t} \mathcal{J}_{t}(\boldsymbol{z}, \bm{\upphi} )},
\end{equation}
where $\boldsymbol{z}$ is the Fourier representation (Eq.~\ref{eq:MFB}), $\bm{\upphi}$ is the harmonic phase representation (Eq.~\ref{eq:harmonic_phase_representation}) subject to the associated bounds. For each iteration the design sensitivities with respect to both entity groups, $\partial \mathcal{J}_{t} / \partial \boldsymbol{z}$ and $\partial \mathcal{J}_{t} / \partial \bm{\upphi}$, are computed through backpropagation (Figure~\ref{backprop})~\cite{paszke2017automatic}.

        \subsection{Data Acquisition Strategy}
As a class of operator learning, IFNO demands a finite collection of data pairs to approximate the joint mapping among the function spaces of interest, e.g., $\mathcal{M}_{AS}:\mathcal{A} \times \mathcal{S} \rightarrow \mathcal{E} $ in our case study, where $\mathcal{A}$, $\mathcal{S}$, $\mathcal{E}$ are associated with a meta-atom array $\chi(x,y)$, modulated phase field $\phi(x,y)$, electric energy intensity $||\mathbf{E}(x,y)||^2$, respectively. A training dataset is structured in the form of
\begin{equation}
\mathcal{D}=\{ (\chi(x,y), \phi(x,y); ||\mathbf{E}||^2(x,y)) | \boldsymbol{z} \in \Theta_A, \bm{\upphi} \in \Theta_S \}     
\end{equation}
where $\chi(x,y)$ is reconstructed from $\boldsymbol{z}$ (Eqs.~\eqref{eq:FMB_1} and \eqref{eq:FMB_thresholding_from_phi}) and $\phi(x,y)$ is specified from $\bm{\upphi}$ (Eq.~\ref{eq:harmonic_phase_representation}).

It has been empirically seen in the corpus of the palette approach, when aiming to build a conventional meta-atom-to-spectrum surrogate of quality, the order of training data typically ranges from $\mathcal{O}(10^3)$ to $\mathcal{O}(10^4)$. Preparing a comparable amount of observational data for both architecture field and phase field could be very resource-intensive. To this end, we implement an efficient data acquisition strategy. In Section~\ref{Data Generation: Meta-Atom Synthesis through Fourier multiclass blending} a 15k-size shape-only dataset $\mathcal{D}_{S}$ born from the six meta-atom families has been prepared. Taking the 36D Fourier representation $\boldsymbol{z}$ as the shape descriptor, we apply shape diversity based sequential acquisition proposed in Lee et al.~\cite{lee2023t}, in order to sequentially identify a 500-size subset with largest shape diversity. Regarding phase, on the other hand, Optimal Latin Hypercube Sampling~\cite{jin2003efficient} is employed to identify 36 space-filling samples, all at once, with the side constraints imposed on the phase variables $\bm{\upphi}$ taken into account. The resulting dataset $\mathcal{D}$ with all the responses available ends up containing 500$\times$36 input instances. We empirically confirmed that $\mathcal{M}_{AS}$ trained on $\mathcal{D}$ exhibits decent predictive performance (Section~\ref{predictive performance}), even with the presence of huge fluctuations of local energy distributions in the output fields (Section~\ref{Modeling: Field-to-Field Modeling through Implicit Fourier Neural Operator}). In case where a more rigorous answer to the question ``How much data?'' is sought, deep active learning~\cite{ren2021survey} may secure more thrifty via directly including a particular ML model of interest into the loop of data acquisition.


\section{Result}

    \subsection{Predictive Performance}
    \label{predictive performance}
Details on the training of $\mathcal{M}_{AS}$ are stated in Appendix \textcolor{blue}{A}. Figure~\ref{Fig4} gives a visual impression on the predictive performance of the trained $\mathcal{M}_{AS}$ with respect to the test shape dataset. The 3$\times$3 plasmonic metasurface array of interest is geometrically periodic, yet able to produce heterogeneous energy distributions when illuminated with the heterogeneously modulated phase distributions. The five randomly selected meta-atom instances $\chi(x,y)$ manifest huge topological freedom, associated with holes, gaps, and organic boundary variations, achievable through the proposed FMB (Section~\ref{Data Generation: Meta-Atom Synthesis through Fourier multiclass blending}). Meanwhile, the phase distributions $\phi(x,y)$ are fully specified by the six phase variables $\bm{\upphi}$ according to Eq.~\ref{eq:harmonic_phase_representation} and passed to $\mathcal{M}_{AS}$ through another input channel. The interplay between meta-atom $\chi$ and incident phase $\phi$ gives rise to a diverse array of energy distribution patterns, as seen in the third column of Figure~\ref{Fig4}. The key challenge herein from modeling perspective is to capture the locally distributed plasmonic energy hotspots, typically exhibiting $\mathcal{O}(10^3)$ stronger energy intensity than the surrounding, formed either along part of the meta-atom boundary or between a gap. For all the cases, the prediction by $\mathcal{M}_{AS}$ shows good agreement with the ground-truth counterpart. More examples with respect to both training and test sets are listed in \B{ Supporting Information} (Figures \textcolor{blue}{S8}-\textcolor{blue}{11} under Section \textcolor{blue}{D}).

\begin{figure}[!htb]
\centering
\includegraphics[width=0.9\textwidth]{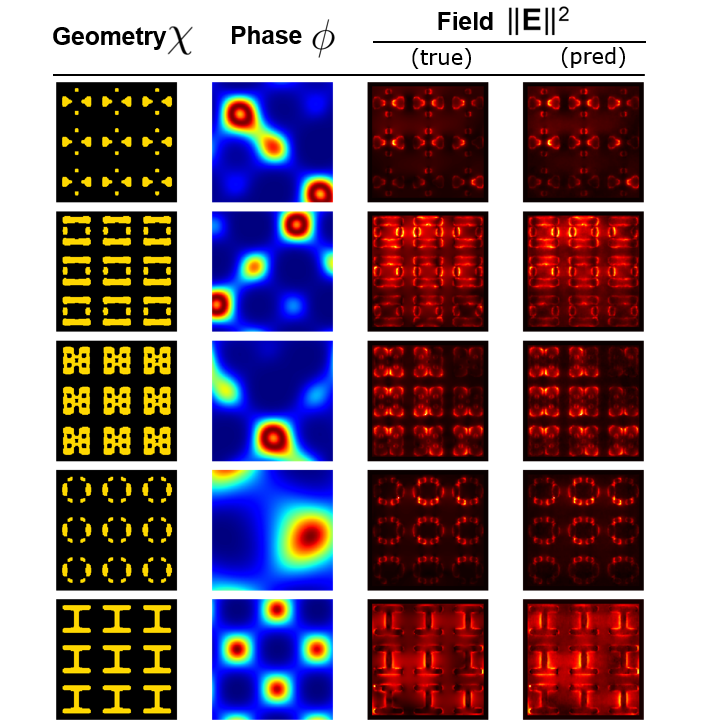}
\caption{Prediction results of $\mathcal{M}_{AS}$ for five randomly selected pairs of meta-atoms and input phase fields from the test dataset.}
\label{Fig4}
\end{figure}
 


    \subsection{Single-Task Concurrent Optimization}    
\label{Single-Task Concurrent Optimization}
Without loss of generality, we will consider the task set $\mathcal{T}=\{1, 2, 3\}$ specified in Section~\ref{phase representation}. As a prior step to the proposed multitask concurrent optimization, single-task concurrent optimization is run for the individual tasks. By doing so we intend to:
\begin{itemize}
    \item corroborate the efficacy of the proposed A-S concurrent optimization. The optimized results with diverse topologies as well as case-specific phase distributions are an indication that the concurrent design is necessary to avoid suboptimality.
    \item identify the upper bounds of $\mathcal{J}_t$ that the following multiobjective optimization can reach for each target pattern. In other words, the values will enable to assess the multiobjective optimization results, in relation to the single objective ones. The resulting objective values $\mathcal{J}_t (t \in \mathcal{T})$ read 3.28, 3.11, 3.16, respectively.
\end{itemize}
For an individual task $n_{rep}=100$ attempts with random initialization were made to mitigate the initial dependence of gradient-based search in the vast joint design space. 
Herein we postulate the task-specific results as referential bounds for the following multitask concurrent optimization.


    \subsection{Multitask Concurrent Optimization}    
\label{Multitask Concurrent Optimization}    
\begin{figure}[!htb]
\centering
\includegraphics[width=0.8\textwidth]{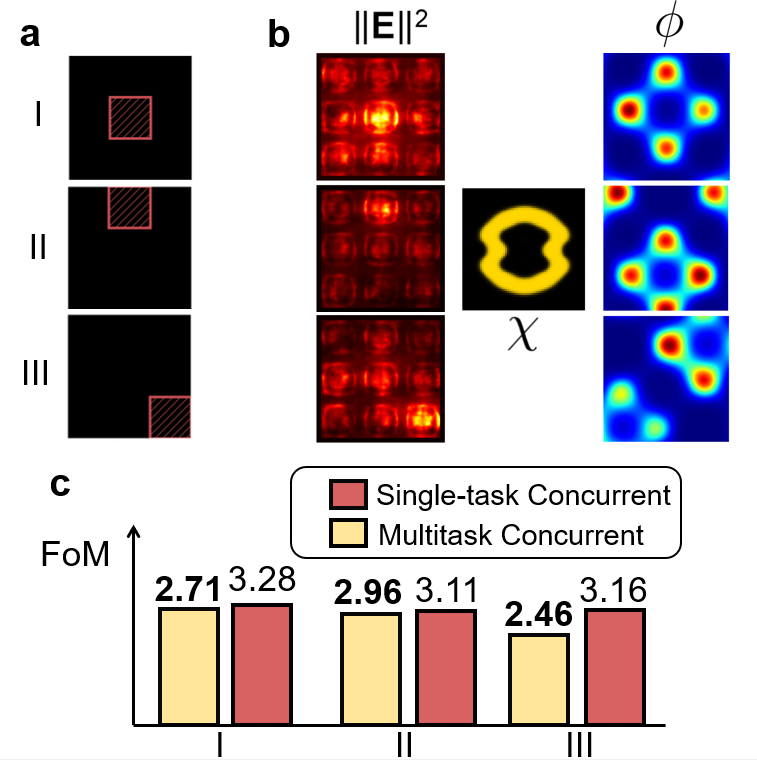}
\caption{A Pareto-optimal solution of the proposed multitask optimization constructed for $\mathcal{T}=\{1, 2, 3\}$. (a) A selected set of target patterns, where a target focusing region is marked with red box. (b) The optimized energy distributions $\{ ||\mathbf{E}||_{1}^2, ||\mathbf{E}||_{2}^2, ||\mathbf{E}||_{3}^2 \}$ that are programmable through the Pareto-optimal meta-atom $\chi^{*}$ paired with the optimized set of task-specific stimuli $\{ \phi^{*} \}=\{\phi_{1}^{*}, \phi_{2}^{*}, \phi_{3}^{*} \}$. (c) The figure-of-merits for each task.}
\label{multitask_concurrent}
\end{figure}

Figure~\ref{multitask_concurrent} illustrates an instance of the Pareto-optimal solutions with respect to the task set $\mathcal{T}=\{1,2,3\}$, which was identified through the proposed multitask concurrent optimization (Section~\ref{Optimization An Inverse Problem Formulation on Multitask Concurrent Optimization of Multifunctional Devices}). The optimized meta-atom $\chi_{{\mathcal{T}}}^{*}$ ends up converging to an inter-class instance. For each target pattern, the phase variable $\bm{\upphi}_{\mathcal{T}}$ is optimized to tailor the resulting energy distribution to be as close as to the target. A qualitative observation on the optimized phase distributions $ \phi_{\mathcal{T}}^{*} $ indicates that the local hills of phase lags tend to suppress energy localization around them. The FoMs with respect to each task reads 2.71, 2.96, and 2.46, respectively. Recalling the single-task optimization result, it seems that this Pareto solution supports a decent performance for Pattern II ($\mathcal{J}_2=2.96$ vs $3.11$), despite the presence of the other two patterns simultaneously taken into account during the optimization. The performance as good as that of the single-task optimization, however, could have come at the cost of the performance gap against the single-task counterpart in Pattern III ($\mathcal{J}_3=2.46$ vs $3.16$). 

Looking into the Pareto solutions all together, we further investigate the trade-off among tasks in the proposed multitask concurrent optimization. The 3D FoM space in Figure~\ref{multitask_concurrent}(a) shows the scatter plot of the 100 optimized solutions $\{ \boldsymbol{z}_k^{*} \}_{k=1}^{n_{rep}}$, each of which started with random initialization on $\boldsymbol{z}$. Denoted with red star, the 15 Pareto solutions form the finite Pareto set $\mathcal{P}_{\theta}$, which is a finite collection of landmarks for the Pareto surface $\mathcal{P}_{\mathbf{J}}$. The ground-truth surface can be approximated as $\hat{\mathcal{P}_{\mathbf{J}}}$ with upon proper surface fitting (yellow region in Figure~\ref{multitask_concurrent}). 2D scatter plots in Figure~\ref{multitask_concurrent}(b)-(d) are projections of Figure~\ref{multitask_concurrent}(a) onto the three principal planes. The projections reveal that the FoMs of interest manifest a trade-off in design of MMM. Without an arbitrary, user-defined importance specified, global optimality across the Pareto solutions would remain ambiguous. 

\begin{figure}[!htb]
\centering
\includegraphics[width=0.9\textwidth]{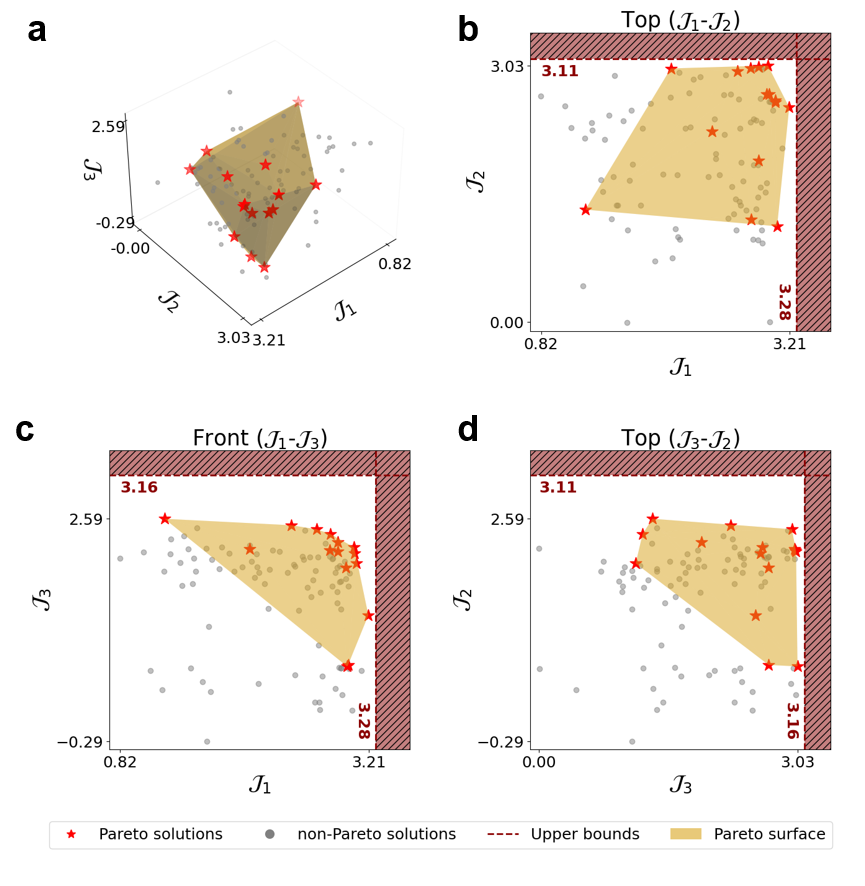}
\caption{
Scattering plots of the Pareto-optimal solutions. The upper bounds for multitask concurrent optimization (red dotted lines) have been identified by the single-task counterpart individually conducted for each task (Section~\ref{Single-Task Concurrent Optimization}). (a) The yellow surface denotes the Pareto surface $\mathcal{P}_{\mathbf{J}}$, approximated as the convex hull of the Pareto set $\mathcal{P}_{\bm{\uptheta}}$. 
(b)-(d) 2D projections of the Pareto surface $\mathcal{P}_{\mathbf{J}}$.
}.
\label{pareto_surface}
\end{figure}

\begin{figure}[!htb]
\centering
\includegraphics[width=0.7\textwidth]{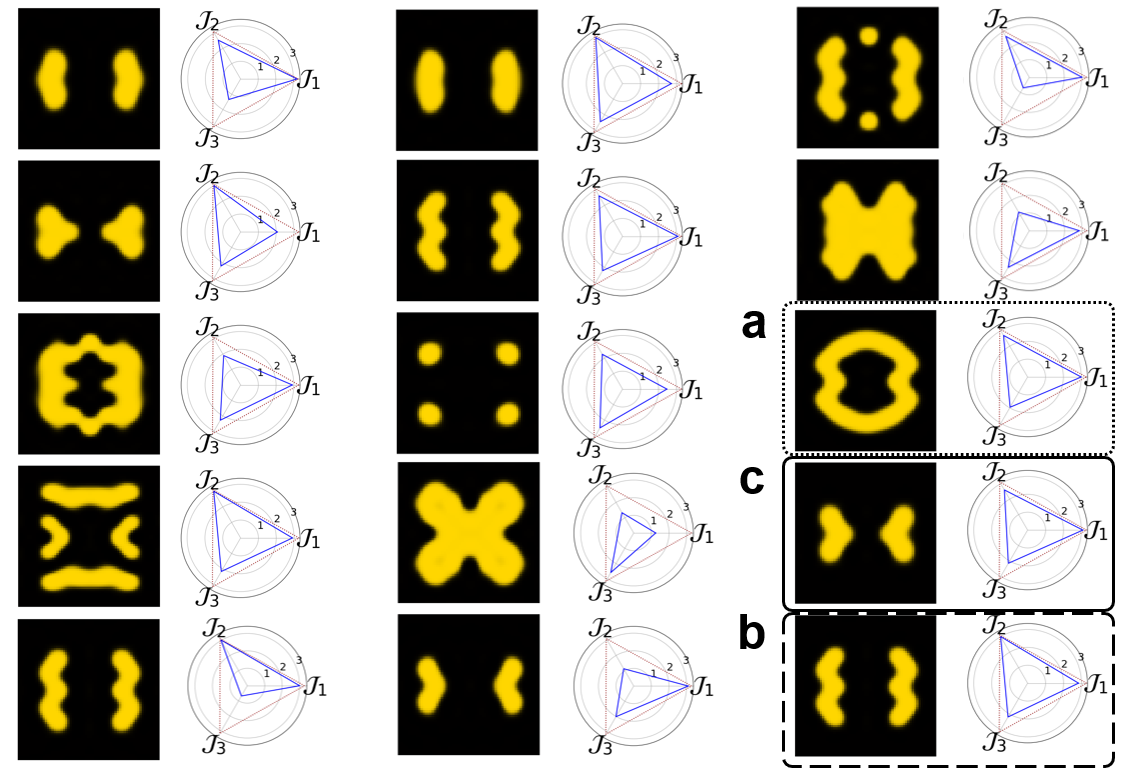}
\caption{Task importance specificity of multifunctional metasurfaces. Optimized meta-atoms in the Pareto set with their radar plots for the FoMs are enumerated. (a)-(c) The best solutions where a hypothetical task importance $(\mathcal{J}_1, \mathcal{J}_2, \mathcal{J}_3$) is given (a)  (1, 1, 1). (b) (2, 1, 1). (c) (5, 1, 4).}
\label{pareto_geometry_task_importance}
\end{figure}

In Figure~\ref{pareto_geometry_task_importance}, all the optimized meta-atoms in the Pareto set are enumerated with the corresponding radar plot of FoMs. By definition, a Pareto solution outperforms another regarding at least one task (Section~\ref{Optimization An Inverse Problem Formulation on Multitask Concurrent Optimization of Multifunctional Devices}). The diverse topology observed in these Pareto-optimal meta-atoms -- arguably blended from bow-tie, ellipse, ring -- is a strong indication of (1) \textit{task importance specificity} in design of multifunctional systems and (2) one-to-many mapping in the inverse problem involving the joint $\mathcal{A}$-$\mathcal{S}$ space, an extension of that in conventional geometry-only inverse problems. Meanwhile, the radar plots show FoM footprints, which seem as diverse as the meta-atoms; this implies that the proposed multitask concurrent optimization can accommodate on-demand task importance, no matter whether it has been declared in advance or yet to be specified. Figure~\ref{pareto_geometry_task_importance}(a)-(c) shows such an example of the latter, where the relative task importance for $\mathcal{T}$ is hypothetically given as $(1, 1, 1)$, $(2, 1, 1)$, $(5, 1, 4)$, respectively. 

So far we have intentionally limited our discussion for the specific task set $\mathcal{T}$. 
We claim that the key observations and findings discussed in this section generalize to other task sets.

    \subsection{Discussion}
    \label{discussion}
\paragraph{Alternatives to the Input Representations}
Through the case study, we have shown the trained neural operator $\mathcal{M}_{AS}$ can be plugged into the Fourier meta-atom representation $\boldsymbol{z}$ and the harmonic phase $\bm{\upphi}$ for inverse design purposes. Without re-training, the model can also be connected to other input representations, as long as the training data well covers the space spanned by the chosen representations. According to the primary concern of inverse design, alternative meta-atom representations include latent representation learned from generative models~\cite{liu2018generative, ma2019}, fabrication-aware representation~\cite{chen2020design, hammond2021photonic, tanriover2022deep}, mixed qualitative-quantitative representation~\cite{so2019designing, wang2021data}, to name a few. Different stimulus representations can also be of interest~\cite{kao2012}. In case the training data does not cover selected input representations well, the neural operator can be re-trained on either (1) new out-of-distribution sparse observations or (2) physics-based residual, as has been postulated by some recent works scrutinizing extrapolation capability of deep neural operators~\cite{zhu2023reliable}.

\paragraph{Model Transparency}
The proposed neural operator based modeling, essentially, is a field-to-field surrogate that offers full-field predictions. By construction, the approach features model transparency: plausibility of the field prediction can be directly inspected with relevant domain knowledge, e.g., strong energy localization forming either between a gap or around meta-atom boundaries. This is a sharp contrast to commonplace black-box models that are often given a direct ``shortcut'' to the output quantity of interest at downstream (e.g., spectra of transmission/phase delay). Despite the practicality and easy implementation, the modeling approach may merely imitate the scratch of underlying light-matter interactions dictated by rich spatiotemporal causal effects. In our case the proposed neural operator combats this issue via encoding long-range, across-meta-atom spatial interactions into the model. Importantly, the advantage comes without extra cost of the data acquisition, compared to that of conventional models fed with full-field simulation data; a possible exception would be the training data computed through analytic approximations that directly give spectral quantities of interest without full-field information computed.

\paragraph{Extending to Geometrically Aperiodic Arrays}
Another assumption regarding the modeling and inverse design presented is that meta-atoms on the array are periodic. We argue that the proposed framework as is has no technical hurdles to impede both modeling and inverse design of geometrically aperiodic arrays. In fact, the proposed framework could rather be an exclusive means to conduct a system-level top-down inverse design all at once, as opposed to the bottom-up camp popular in literature, where a massive meta-atom library is prepared --- often under negligible meta-atom coupling -- and then simply tiled in the array during the system-level design. A conceptual extension of the meta-atom library based approach has been proposed by An et al.~\cite{an2022deep}, where a supercell library is prepared to compensate for the deviations of response spectra under near-field coupling effects. Even with the extension, however, the surrogate does not offer field predictions and the supercell-based inverse design is intrinsically restricted to the class of design problems where required distributions of transmission/phase delay have been identified (e.g., meta-lenses). Such cases can be handled by our proposed framework. 
We envision that the extension of the proposed framework to aperiodic meta-atom arrays would be feasible with (1) intelligent data acquisition strategies redefining shape diversity for supercells -- not for unit cells -- as a core pillar and (2) gradient-based search with a large number of randomly initialized replicates to deal with the extra dimensionality incurred in the architecture space. 

\paragraph{Other Optimization Methods}
Throughout this article the design sensitivities of interest, e.g., $\partial \mathcal{J} / \partial \boldsymbol{z} $ and $\partial \mathcal{J} / \partial \bm{\upphi} $, are assumed to be accessible through Automatic Differentiation. Augenstein et al.~\cite{augenstein2023neural} articulated that the combination of a data-driven surrogate and gradient-based search allows to offset the cost for data acquisition and model construction. Echoing the claim, we have also harnessed gradient-based, iterative design update, summarized in Eqs.~\ref{eq: design update for stimulus} and \ref{eq: design update for architecture}, with a number of random initial starting points. Nevertheless, we reflect upon the claim via postulating some scenarios, beyond our case study, that could make the claim refutable: (i) the design sensitivity is not reliable enough to be directly used for design search and (ii) the predictive performance of associated ML model is not accurate enough, hence for concurrent optimization it seems more reasonable to use the ground-truth solver with thrifty sequential sampling. If such is the case, the proposed formalism in Eq.~\ref{eq:multiple task loss} may be better addressed through efficient global optimization algorithms, e.g., Bayesian optimization~\cite{snoek2012}, provided that the scalability issue with respect to input dimensionality can be somehow resolved. Regarding multitask/multiobjective Bayesian optimization, readers are referred to some foundational works~\cite{swersky2013multi-task}. In the meantime, the widely-used metaheuristic optimization~\cite{halim2021performance}, e.g., generic algorithms and particle swarm optimization, in general, seems not effective for inverse design of MMM in that: (1) the vast, high-dimensional design space jointly formed by architecture and stimulus and (2) many optimization hyperparameters that are critical for search performance thus supposed to be exhaustively fine-tuned.

\section{Conclusion}
In this paper, we aim to address three grand challenges associated with the design of multifunctional metamaterials involving heterogenous fields, namely, (1) vast design space jointly formed by architecture, stimulus, and optionally material, (2) a prevalent trade-off across multiple functionalities of interest, and (3) a lack of standardized inverse problem formulations on multifunctional metamaterials and solution procedures thereof. To overcome the limitations of the palette approach that assumes scale separation, we presented a data-driven design framework that can streamline the inverse design of multifunctional metamaterials whose functionality/operating conditions feature heterogeneous fields. The framework interlocks three methodological pillars: 
\begin{itemize}
    \item Implicit Fourier neural operator, which serves as a field-to-field surrogate from a pair of architecture-stimulus fields to the corresponding high-dimensional, possibly heterogeneous physical fields.  
    \item A standard formulation of the inverse problem on a class of multifunctional metamaterials, where a system is open to both architecture and stimulus, and multiple target functionalities often subject to a trade-off. We also propose a principled, gradient-based solution procedure involving the joint architecture-stimulus space, with Pareto-optimality of the multifunctionality addressed. 
    \item Fourier multiclass blending as a new data generation scheme. It facilitates to accommodate domain knowledge, produce inter-class, quasi-free meta-atoms with smooth topological transition, and features training-free dimension reduction and built-in reconstruction. 
\end{itemize}
By seamlessly integrating the three pillars, we demonstrated our approach to the inverse design on plasmonic metasurfaces whose field distribution can be dynamically programmable through spatial light modulators. Our proposed approach addresses simultaneously the aforementioned three grand challenges. The prediction results of the proposed implicit Fourier neural operator demonstrated the satisfactory predictive performances over heterogeneous energy fields, featuring plasmonic hotspots with an energy intensity on the order of $10^3 \times$ stronger than the surrounding, with respect to virtually infinite field-type pairs of quasi-free supercell and incident phase. The optimization results corroborated that the proposed framework can automatically identify a Pareto set of meta-atom and incident phase. Looking into the Pareto-optimal solutions, we reported huge diversity regarding both meta-atom topology and figure-of-merit profiles, which could handle both mutable task importance specificity and the one-to-many mapping of inverse design. In addition to the technical contributions and observations centered on the case study, we also articulated three crucial advantages that the proposed framework offers by construction, namely (1) direct connectivity to alternative combinations of input representations, (2) model transparency that allows a physics-based sanity check, and (3) sample-/parameter-efficiency.  In a broader sense, we shared our perspectives on feasible extensions of our framework to accommodate geometrically aperiodic arrays, and other optimization algorithms as well. Collecting evidence to support these claims with the design scenarios not directly covered in our case study would be our future work. We believe that this design framework, succeeding to the palette approach, qualifies as an important step toward next-generation data-driven design for multiscale architectures.


\medskip
\textbf{Acknowledgements} \par 
W. Chen and D. Lee appreciate the support by the NSF BRITE Fellow program (CMMI 2227641), the NSF CSSI program (OAC 1835782), and the Northwestern McCormick Catalyst Award. Y. Yu and L. Zhang would like to acknowledge support by the NSF Award (DMS-1753031) and the AFOSR grant (FA9550-22-1-0197). Portions of this research were conducted on Lehigh University's Research Computing infrastructure partially supported by NSF Award (2019035).

\medskip


\section*{Appendix A. Training Details of the Proposed Implicit Fourier Neural Operator $\mathcal{M_{AS}}$}
\label{SI_Training}

\paragraph{Scheduling}
When training a machine learning model based on stochastic gradient descent, a source of noise is introduced due to the random sampling of training instances, and does not vanish even when we arrive at a minimum~\cite{goodfellow2016deep}. A workaround for better convergence is to employ a learning rate schedule which gradually decreases the learning rate between epochs as the training progresses. Following this, in this work, we employ a learning rate schedule which gradually decreases the learning rate between epochs as the training progresses. During the network training the learning rate $lr^{(i)}$ for each iteration $(i)$ is recursively updated according to
\begin{equation}
lr^{(i)} = lr_{0}  {\gamma}^{\left\lfloor \frac{(i)}{n_{step}} \right\rfloor},
\label{scheduling}
\end{equation}
where $lr_{0}$ is the initial learning rate set as $lr_{0}=lr=0.01$ in our case; $n_{step}$ is the update period set as 100 in our training. The scheduling is incorported into Adam optimizer~\cite{kingma2014adam}.

\paragraph{Training}
\label{Model Training}

\begin{figure}[t!]
\centering
\includegraphics[width=0.9\textwidth]{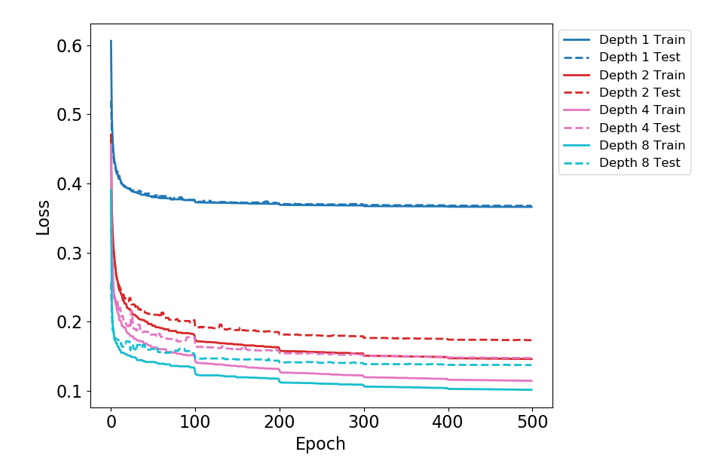}
\caption{Training history of $\mathcal{M}_{AS}$ for different layer depths.}
\label{SI_history_depth}
\end{figure}

Figure~\ref{SI_history_depth} displays the training results with regard to different depths. All the results show acceptable differences of loss between training and test data. We conducted parameter study on depth and empirically chose depth=8 considering (1) the balance between model complexity and predictive performance and (2) the gap between $\mathcal{L}_{train}$ and $\mathcal{L}_{test}$. The discrete jumps in each training run are involved with the scheduling, following Eq.~\ref{scheduling}. The proposed IFNO $\mathcal{M}_{AS}$ is open for further performance improvement upon hyperparameter study on model complexity as well as training parameters, using either conventional grid search or Bayesian optimization~\cite{shahriari2015taking, snoek2012practical}.

\section*{Appendix B. The Wave Analysis Simulation}
\label{Wave_Anaylsis}
\begin{figure}[t!]
\centering
\includegraphics[width=0.8\textwidth]
{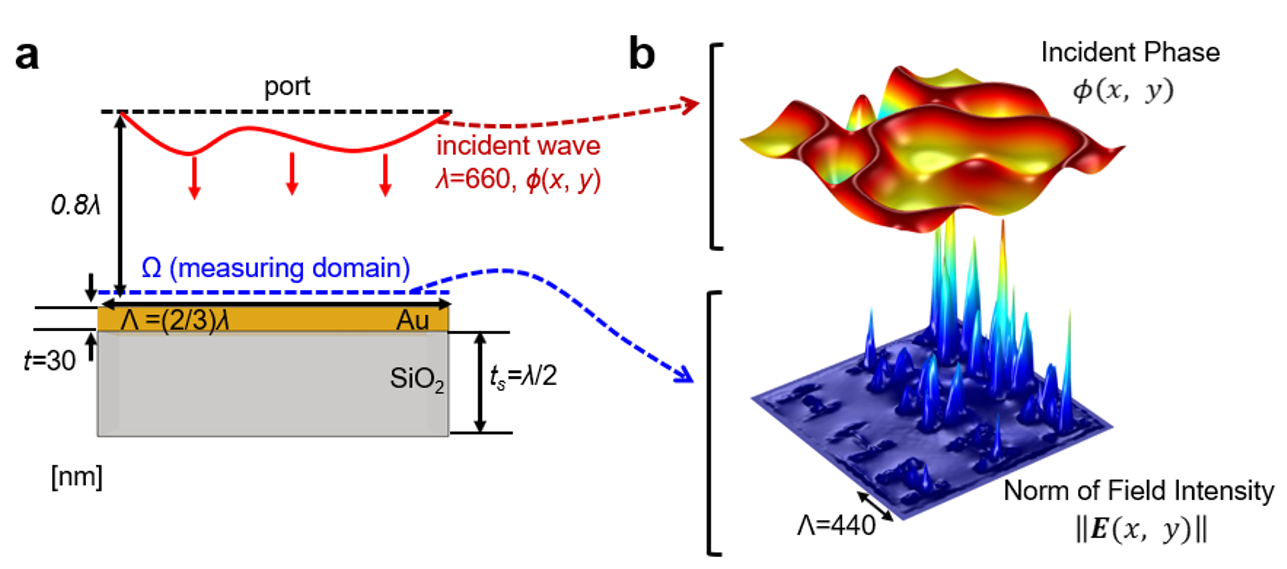}
\caption{A schematic of wave analysis. (a)  A side view of the whole plasmonic metasurface system. (b) Virtual profiles of a modulate incident phase and the resulting norm of electric field intensity.}
\label{fig:wave_analysis}
\end{figure}

A visual illustration of the wave analysis is in Figure~\ref{fig:wave_analysis}. The wave of incidence has a wavelength of $660$ nm (or a frequency of $454$ THz) and is polarized in the $x$-direction. The input phase is adjusted by the six design variables from our suggested phase representation. The periodicity is set at $440$ nm, which is two-thirds of the wavelength. The unit nanoantenna, composed of gold with a permittivity of $\varepsilon=-13.682+1.3056i$ at frequency $f$, forms an array through the periodic tessellation of $30$-nm-thick unit cells. This array is set on a cuboid of SiO$_2$, which is half the wavelength thick and has a permittivity of $\varepsilon=3.75$. Perfectly matched layers surround the entire analysis domain to minimize boundary reflections. Frequency Domain in The RF Module of COMSOL Multiphysics~\cite{comsol2021} was used for the full-wave analysis. 

\section*{Appendix C. Fourier Multiclass Blending}
\label{sec:FMB}

\begin{figure}[H]
\centering
\includegraphics[width=0.7\textwidth]{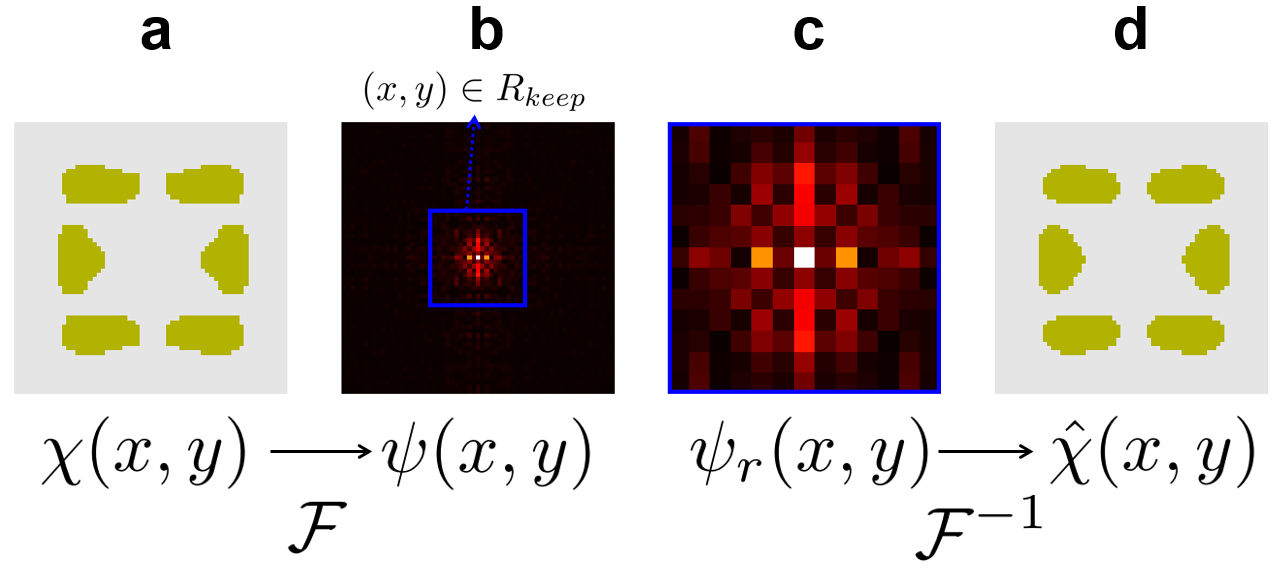}
\caption{A visual pipeline from a binary meta-atom to the Fourier representation. (a) A 2D binary image. (b) The Fourier Transform of the given image. (c) The sparse representation after cropping with padding depth $d$. (d) The reconstructed binary image with inverse Fourier Transform.}
\label{fig:FT representation pipeline}
\end{figure}

\begin{figure}[htb!]
\centering
\includegraphics[width=0.7\textwidth]{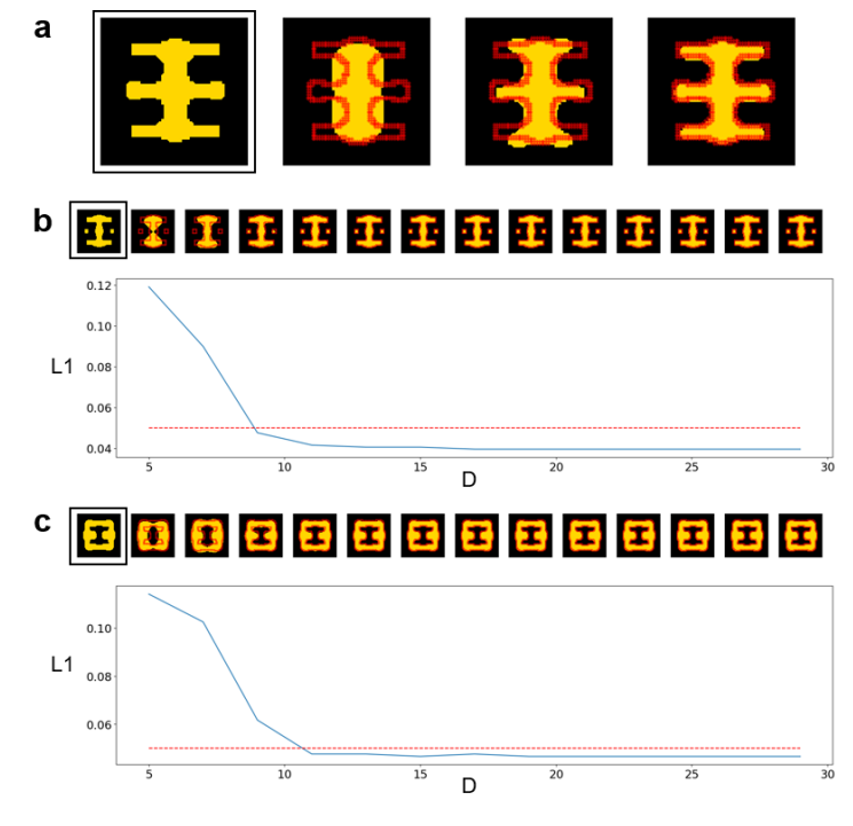}
\caption{
Built-in reconstruction with the Fourier representation. All the black boxes denote the original binary image $\chi(x,y)$, while all the red lines denote the boundaries of it. (a) Reconstructed images $\chi_r(x,y)$ of the given unit cell for $d=5, 7, 9$, respectively. (b) L1 error of reconstruction as a function of dpeth $d$. The red dotted line denotes 5\% error. (c) L1 error of reconstruction of another unit cell.
}
\label{fig:SI_FMB_feature_dimension}
\end{figure}

\begin{figure}[t!]
\centering
\includegraphics[width=0.6\textwidth]{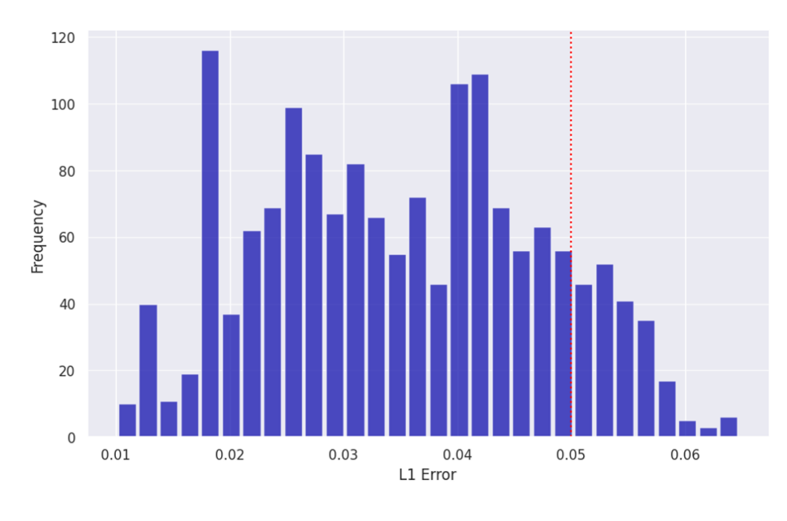}
\caption{
L1 error distribution of reconstruction under $d=11$ and a error threshold 0.05. Approximately 87.2\% shows less error than the threshold.
}
\label{SI_FMB_error_historgram}
\end{figure}

\begin{figure}[t!]
\centering
\includegraphics[width=1\textwidth]{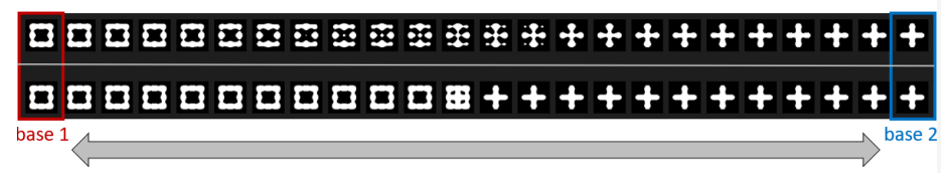}
\caption{
Visual comparison of linear traversal between (top) the proposed blending (FMB) and (bottom) blending in the image space.
}
\label{SI_FMB_smooth_comparison}
\end{figure}

Our implementation procedure of Fourier multiclass blending builds on the topological encoding procedure proposed by Liu et al.~\cite{liu2020topological}. Upon proper translations, high-frequency components of the Fourier representation are located along the boundary of the frequency image $\psi(x,y)$. $R_{keep}$ denotes the internal region to be preserved as is, while the complementary region will be zeroed out. $R_{keep}$ is specified through the depth $d$ that represents the thickness of padding (Figure~\ref{fig:FT representation pipeline}). The resulting dimension of the sparse representation becomes $(N-2d)^2$. For instance, if resolution $N$ and depth $d$ happen to be 64 and 30, respectively, the corresponding dimension of $\psi_r(x,y)$ would reduce to $(64-2\times30)^2=16$. Figure~\ref{fig:SI_FMB_feature_dimension} gives an example of reconstruction errors with respect to $d$ for a couple of binary images. By construction, the Fourier transform produces the frequency map that inherits some types of symmetries in the image space. For example, given a binary image $\chi(x,y)$ with four-fold symmetry, the corresponding sparse representation $\psi_r(x,y)$ also holds four-fold symmetry, thus is reducible to one quadrant. The resulting dimensionality reduces from $N \times N$ to $(N-2d)^2/4$. Revisiting the above-mentioned case where $N=64$ and $d=30$, this results in dimension reduction from $64^2$D to $4$D. Based on our empirical observation on the L1 error distribution, visualized in Figure~\ref{SI_FMB_error_historgram}, $d=11$ was chosen as the padding depth. This results in the feature matrix of $\{(11+1)/2\}^2=36$D. For notational simplicity, the flattened vector of the matrix, which would be called Fourier feature, is to be used. This is the final form of the Fourier representation $\boldsymbol{z}$ used for Fourier multiclass blending in the main body. So far the description on the encoding procedure has assumed its deployment to 2D images in $\mathbb{R}^2$ space. The procedure trivially generalizes to $\mathbb{R}^3$ space.

\begin{figure}[hbt!]
\centering
\includegraphics[width=0.9\textwidth]{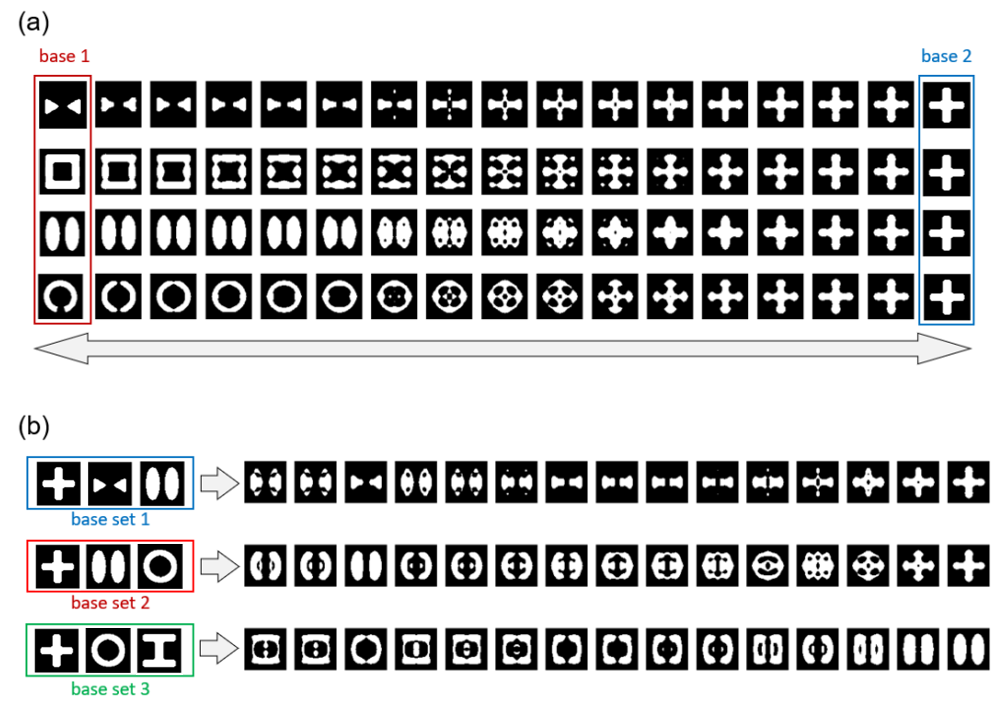}
\caption{
Additional examples of (a) two-class blending, ordered monotonically hence can be seen as linear traversal and (b) three-class blending, displayed without a specific order.
}
\label{SI_FMB_example_instances}
\end{figure}



\clearpage

\section*{Appendix D. Prediction Results of $\mathcal{M}_{AS}$: Additional Examples}
\label{More Examples of Prediction Results}

\begin{figure}[h!]
\centering
\includegraphics[width=.9\textwidth]{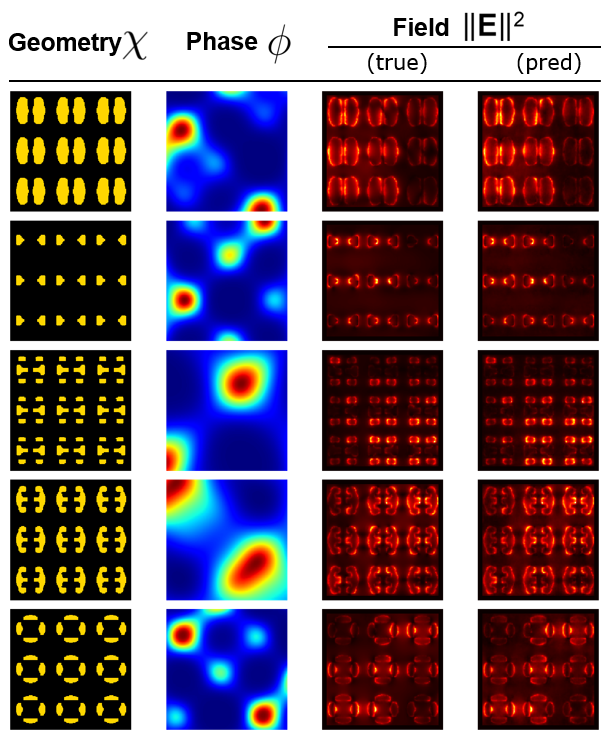}
\caption{
Prediction results of $\mathcal{M}_{AS}$ for five randomly selected from the training samples (1/2). All scale bars for each column has an arbitrary unit.
}
\label{SI_train_1}
\end{figure}

\begin{figure}[H]
\centering
\includegraphics[width=.9\textwidth]{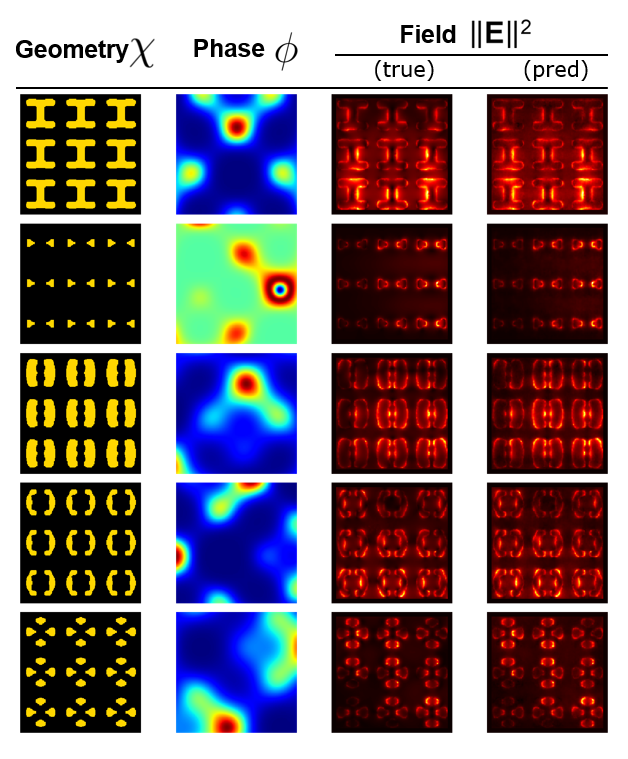}
\caption{
Prediction results of $\mathcal{M}_{AS}$ for five randomly selected from the training samples (2/2). All scale bars for each column has an arbitrary unit.
}
\label{SI_train_2}
\end{figure}

\begin{figure}[H]
\centering
\includegraphics[width=.9\textwidth]{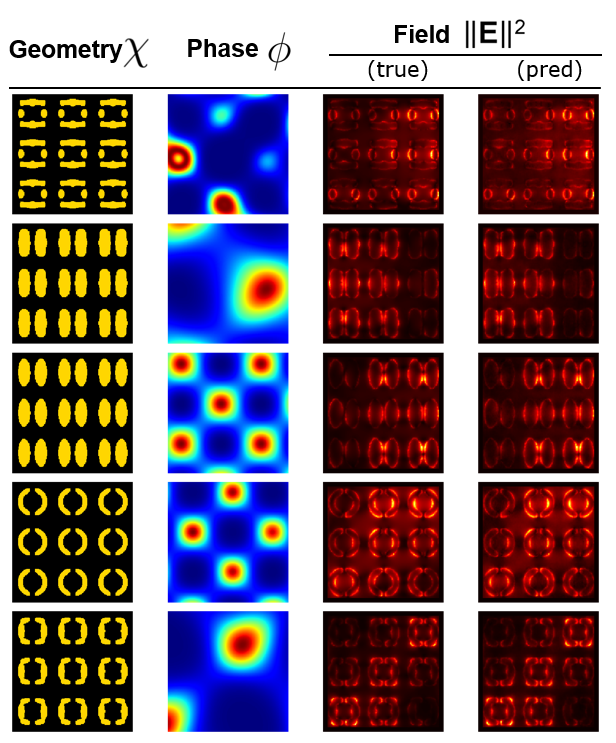}
\caption{
Prediction results of $\mathcal{M}_{AS}$ for five randomly selected from the test samples (1/2). All scale bars for each column has an arbitrary unit.
}
\label{SI_test_1}
\end{figure}

\begin{figure}[H]
\centering
\includegraphics[width=.9\textwidth]{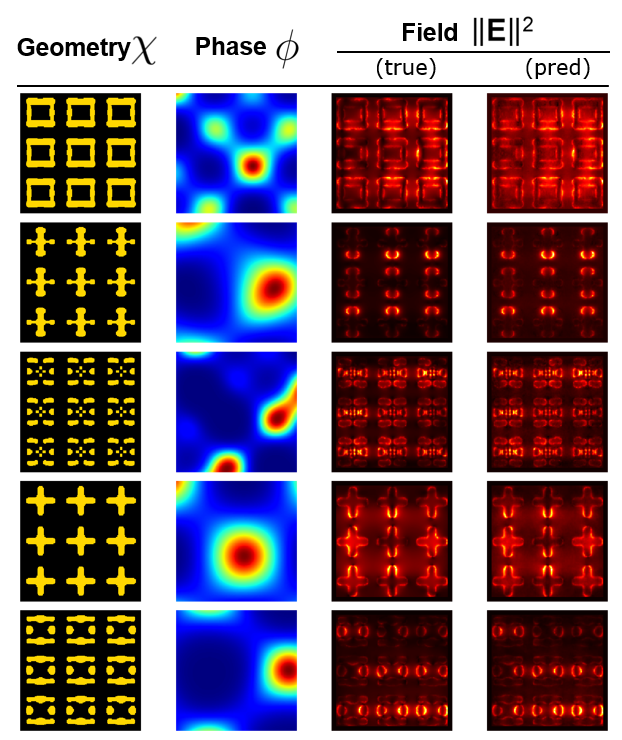}
\caption{
Prediction results of $\mathcal{M}_{AS}$ for five randomly selected from the test samples (2/2). All scale bars for each column has an arbitrary unit.
}
\label{SI_test_2}
\end{figure}

\bibliography{mainarXiv}

\begin{thebibliography}{100}

\bibitem{pendry2000negative}
John~Brian Pendry.
\newblock Negative refraction makes a perfect lens.
\newblock {\em Physical review letters}, 85(18):3966, 2000.

\bibitem{yu2018mechanical}
Xianglong Yu, Ji~Zhou, Haiyi Liang, Zhengyi Jiang, and Lingling Wu.
\newblock Mechanical metamaterials associated with stiffness, rigidity and compressibility: A brief review.
\newblock {\em Progress in Materials Science}, 94:114--173, 2018.

\bibitem{lincoln2019multifunctional}
Reece~L Lincoln, Fabrizio Scarpa, Valeska~P Ting, and Richard~S Trask.
\newblock Multifunctional composites: A metamaterial perspective.
\newblock {\em Multifunctional Materials}, 2(4):043001, 2019.

\bibitem{wu2019mechanical}
Wenwang Wu, Wenxia Hu, Guian Qian, Haitao Liao, Xiaoying Xu, and Filippo Berto.
\newblock Mechanical design and multifunctional applications of chiral mechanical metamaterials: A review.
\newblock {\em Materials \& design}, 180:107950, 2019.

\bibitem{yuan2021recent}
Xujin Yuan, Mingji Chen, Yin Yao, Xiaogang Guo, Yixing Huang, Zhilong Peng, Baosheng Xu, Bowen Lv, Ran Tao, Shenyu Duan, et~al.
\newblock Recent progress in the design and fabrication of multifunctional structures based on metamaterials.
\newblock {\em Current Opinion in Solid State and Materials Science}, 25(1):100883, 2021.

\bibitem{deng2018multifunctional}
Zi-Lan Deng, Yaoyu Cao, Xiangping Li, and Guo~Ping Wang.
\newblock Multifunctional metasurface: from extraordinary optical transmission to extraordinary optical diffraction in a single structure.
\newblock {\em Photonics Research}, 6(5):443--450, 2018.

\bibitem{wu2019symmetry}
Shuai Wu, Qiji Ze, Rundong Zhang, Nan Hu, Yang Cheng, Fengyuan Yang, and Ruike Zhao.
\newblock Symmetry-breaking actuation mechanism for soft robotics and active metamaterials.
\newblock {\em ACS applied materials \& interfaces}, 11(44):41649--41658, 2019.

\bibitem{li2019inverse}
Junyu Li, Li~Bao, Shun Jiang, Qiushi Guo, Dehui Xu, Bin Xiong, Guangzu Zhang, and Fei Yi.
\newblock Inverse design of multifunctional plasmonic metamaterial absorbers for infrared polarimetric imaging.
\newblock {\em Optics express}, 27(6):8375--8386, 2019.

\bibitem{al2019multifunctional}
Oraib Al-Ketan and Rashid~K Abu Al-Rub.
\newblock Multifunctional mechanical metamaterials based on triply periodic minimal surface lattices.
\newblock {\em Advanced Engineering Materials}, 21(10):1900524, 2019.

\bibitem{chen2020metasurface}
Shuqi Chen, Wenwei Liu, Zhancheng Li, Hua Cheng, and Jianguo Tian.
\newblock Metasurface-empowered optical multiplexing and multifunction.
\newblock {\em Advanced Materials}, 32(3):1805912, 2020.

\bibitem{ze2020magnetic}
Qiji Ze, Xiao Kuang, Shuai Wu, Janet Wong, S~Macrae Montgomery, Rundong Zhang, Joshua~M Kovitz, Fengyuan Yang, H~Jerry Qi, and Ruike Zhao.
\newblock Magnetic shape memory polymers with integrated multifunctional shape manipulation.
\newblock {\em Advanced Materials}, 32(4):1906657, 2020.

\bibitem{tao2020multifunctional}
Hongcheng Tao and James Gibert.
\newblock Multifunctional mechanical metamaterials with embedded triboelectric nanogenerators.
\newblock {\em Advanced Functional Materials}, 30(23):2001720, 2020.

\bibitem{shirmanesh2020electro}
Ghazaleh~Kafaie Shirmanesh, Ruzan Sokhoyan, Pin~Chieh Wu, and Harry~A Atwater.
\newblock Electro-optically tunable multifunctional metasurfaces.
\newblock {\em ACS nano}, 14(6):6912--6920, 2020.

\bibitem{an2021multifunctional}
Sensong An, Bowen Zheng, Hong Tang, Mikhail~Y Shalaginov, Li~Zhou, Hang Li, Myungkoo Kang, Kathleen~A Richardson, Tian Gu, Juejun Hu, et~al.
\newblock Multifunctional metasurface design with a generative adversarial network.
\newblock {\em Advanced Optical Materials}, 9(5):2001433, 2021.

\bibitem{liu2021multifunctional}
Mingze Liu, Wenqi Zhu, Pengcheng Huo, Lei Feng, Maowen Song, Cheng Zhang, Lu~Chen, Henri~J Lezec, Yanqing Lu, Amit Agrawal, et~al.
\newblock Multifunctional metasurfaces enabled by simultaneous and independent control of phase and amplitude for orthogonal polarization states.
\newblock {\em Light: Science \& Applications}, 10(1):107, 2021.

\bibitem{askari2020additive}
Meisam Askari, David~A Hutchins, Peter~J Thomas, Lorenzo Astolfi, Richard~L Watson, Meisam Abdi, Marco Ricci, Stefano Laureti, Luzhen Nie, Steven Freear, et~al.
\newblock Additive manufacturing of metamaterials: A review.
\newblock {\em Additive Manufacturing}, 36:101562, 2020.

\bibitem{bi2021all}
Ke~Bi, Qingmin Wang, Jianchun Xu, Lihao Chen, Chuwen Lan, and Ming Lei.
\newblock All-dielectric metamaterial fabrication techniques.
\newblock {\em Advanced Optical Materials}, 9(1):2001474, 2021.

\bibitem{yoon2016challenges}
Gwanho Yoon, Inki Kim, and Junsuk Rho.
\newblock Challenges in fabrication towards realization of practical metamaterials.
\newblock {\em Microelectronic Engineering}, 163:7--20, 2016.

\bibitem{levchenko2018hierarchical}
Igor Levchenko, Kateryna Bazaka, Michael Keidar, Shuyan Xu, and Jinghua Fang.
\newblock Hierarchical multicomponent inorganic metamaterials: intrinsically driven self-assembly at the nanoscale.
\newblock {\em Advanced Materials}, 30(2):1702226, 2018.

\bibitem{lichade2021hierarchical}
Ketki~M Lichade, Yizhou Jiang, and Yayue Pan.
\newblock Hierarchical nano/micro-structured surfaces with high surface area/volume ratios.
\newblock {\em Journal of Manufacturing Science and Engineering}, 143(8):081002, 2021.

\bibitem{dolar2023interpretable}
Tuba Dolar, Doksoo Lee, and Wei Chen.
\newblock Interpretable neural network analyses for understanding complex physical interactions in engineering design.
\newblock In {\em International Design Engineering Technical Conferences and Computers and Information in Engineering Conference}, volume 87301, page V03AT03A021. American Society of Mechanical Engineers, 2023.

\bibitem{lee2023data}
Doksoo Lee, Wei~Wayne Chen, Liwei Wang, Yu-Chin Chan, and Wei Chen.
\newblock Data-driven design for metamaterials and multiscale systems: A review.
\newblock {\em arXiv preprint arXiv:2307.05506}, 2023.

\bibitem{So2020DeepNanophotonics}
Sunae So, Trevon Badloe, Jaebum Noh, Junsuk Rho, and Jorge Bravo-Abad.
\newblock {Deep learning enabled inverse design in nanophotonics}.
\newblock {\em Nanophotonics}, 9(5):1041--1057, 2020.

\bibitem{Regenwetter2021DeepReview}
Lyle Regenwetter, Amin~Heyrani Nobari, and Faez Ahmed.
\newblock Deep generative models in engineering design: A review.
\newblock {\em Journal of Mechanical Design}, 144(7):071704, 2022.

\bibitem{Kumar2021WhatMechanics}
Siddhant Kumar and Dennis~M Kochmann.
\newblock What machine learning can do for computational solid mechanics.
\newblock In {\em Current trends and open problems in computational mechanics}, pages 275--285. Springer, 2022.

\bibitem{jin2022intelligent}
Yabin Jin, Liangshu He, Zhihui Wen, Bohayra Mortazavi, Hongwei Guo, Daniel Torrent, Bahram Djafari-Rouhani, Timon Rabczuk, Xiaoying Zhuang, and Yan Li.
\newblock Intelligent on-demand design of phononic metamaterials.
\newblock {\em Nanophotonics}, 11(3):439--460, 2022.

\bibitem{woldseth2022use}
Rebekka~V Woldseth, Niels Aage, J~Andreas B{\ae}rentzen, and Ole Sigmund.
\newblock On the use of artificial neural networks in topology optimisation.
\newblock {\em Structural and Multidisciplinary Optimization}, 65(10):1--36, 2022.

\bibitem{so2022revisiting}
Sunae So, Jungho Mun, Junghyun Park, and Junsuk Rho.
\newblock Revisiting the design strategies for metasurfaces: Fundamental physics, optimization, and beyond.
\newblock {\em Advanced Materials}, page 2206399, 2022.

\bibitem{liu2023deep}
Chen-Xu Liu and Gui-Lan Yu.
\newblock Deep learning for the design of phononic crystals and elastic metamaterials.
\newblock {\em Journal of Computational Design and Engineering}, 10(2):602--614, 2023.

\bibitem{zheng2023deep}
Xiaoyang Zheng, Xubo Zhang, Ta-Te Chen, and Ikumu Watanabe.
\newblock Deep learning in mechanical metamaterials: From prediction and generation to inverse design.
\newblock {\em Advanced Materials}, page 2302530, 2023.

\bibitem{cybenko1989approximation}
George Cybenko.
\newblock Approximation by superpositions of a sigmoidal function.
\newblock {\em Mathematics of control, signals and systems}, 2(4):303--314, 1989.

\bibitem{chen1995universal}
Tianping Chen and Hong Chen.
\newblock Universal approximation to nonlinear operators by neural networks with arbitrary activation functions and its application to dynamical systems.
\newblock {\em IEEE transactions on neural networks}, 6(4):911--917, 1995.

\bibitem{lanthaler2023nonlocal}
Samuel Lanthaler, Zongyi Li, and Andrew~M Stuart.
\newblock The nonlocal neural operator: Universal approximation.
\newblock {\em arXiv preprint arXiv:2304.13221}, 2023.

\bibitem{Zhu2017}
Bo~Zhu, Mélina Skouras, Desai Chen, and Wojciech Matusik.
\newblock {Two-scale topology optimization with microstructures}.
\newblock {\em ACM Transactions on Graphics}, 2017.

\bibitem{schumacher2015microstructures}
Christian Schumacher, Bernd Bickel, Jan Rys, Steve Marschner, Chiara Daraio, and Markus Gross.
\newblock {Microstructures to control elasticity in 3D printing}.
\newblock {\em ACM Transactions on Graphics (TOG)}, 34(4):1--13, 2015.

\bibitem{panetta2015elastic}
Julian Panetta, Qingnan Zhou, Luigi Malomo, Nico Pietroni, Paolo Cignoni, and Denis Zorin.
\newblock Elastic textures for additive fabrication.
\newblock {\em ACM Transactions on Graphics (TOG)}, 34(4):1--12, 2015.

\bibitem{malkiel2018plasmonic}
Itzik Malkiel, Michael Mrejen, Achiya Nagler, Uri Arieli, Lior Wolf, and Haim Suchowski.
\newblock Plasmonic nanostructure design and characterization via deep learning.
\newblock {\em Light: Science \& Applications}, 7(1):1--8, 2018.

\bibitem{ma2019}
W.~Ma, F.~Cheng, Y.~Xu, Q.~Wen, and Y.~Liu.
\newblock Probabilistic representation and inverse design of metamaterials based on a deep generative model with semi-supervised learning strategy.
\newblock {\em Adv. Mater.}, 31(35):1–9, 2019.

\bibitem{liu2018generative}
Z.~Liu, D.~Zhu, S.~P. Rodrigues, K.~T. Lee, and W.~Cai.
\newblock Generative model for the inverse design of metasurfaces.
\newblock {\em Nano Lett.}, 18(10):6570--6576, 2018.

\bibitem{so2019designing}
Sunae So and Junsuk Rho.
\newblock Designing nanophotonic structures using conditional deep convolutional generative adversarial networks.
\newblock {\em Nanophotonics}, 8(7):1255--1261, 2019.

\bibitem{liu2022growth}
Ke~Liu, Rachel Sun, and Chiara Daraio.
\newblock Growth rules for irregular architected materials with programmable properties.
\newblock {\em Science}, 377(6609):975--981, 2022.

\bibitem{moharam1995formulation}
MG~Moharam, Eric~B Grann, Drew~A Pommet, and TK~Gaylord.
\newblock Formulation for stable and efficient implementation of the rigorous coupled-wave analysis of binary gratings.
\newblock {\em JOSA a}, 12(5):1068--1076, 1995.

\bibitem{taflove2005computational}
Allen Taflove, Susan~C Hagness, and Melinda Piket-May.
\newblock Computational electromagnetics: the finite-difference time-domain method.
\newblock {\em The Electrical Engineering Handbook}, 3(629-670):15, 2005.

\bibitem{jin2015finite}
Jian-Ming Jin.
\newblock {\em The finite element method in electromagnetics}.
\newblock John Wiley \& Sons, 2015.

\bibitem{kwon2018nonlocal}
Hoyeong Kwon, Dimitrios Sounas, Andrea Cordaro, Albert Polman, and Andrea Al{\`u}.
\newblock Nonlocal metasurfaces for optical signal processing.
\newblock {\em Physical review letters}, 121(17):173004, 2018.

\bibitem{overvig2020multifunctional}
Adam~C Overvig, Stephanie~C Malek, and Nanfang Yu.
\newblock Multifunctional nonlocal metasurfaces.
\newblock {\em Physical Review Letters}, 125(1):017402, 2020.

\bibitem{capers2021designing}
James~R Capers, Stephen~J Boyes, Alastair~P Hibbins, and Simon~AR Horsley.
\newblock Designing the collective non-local responses of metasurfaces.
\newblock {\em Communications Physics}, 4(1):209, 2021.

\bibitem{an2022deep}
Sensong An, Bowen Zheng, Mikhail~Y Shalaginov, Hong Tang, Hang Li, Li~Zhou, Yunxi Dong, Mohammad Haerinia, Anuradha~Murthy Agarwal, Clara Rivero-Baleine, et~al.
\newblock Deep convolutional neural networks to predict mutual coupling effects in metasurfaces.
\newblock {\em Advanced Optical Materials}, 10(3):2102113, 2022.

\bibitem{noh2022reconfigurable}
Jaebum Noh, Yong-Hyun Nam, Sun-Gyu Lee, In-Gon Lee, Yongjune Kim, Jeong-Hae Lee, and Junsuk Rho.
\newblock Reconfigurable reflective metasurface reinforced by optimizing mutual coupling based on a deep neural network.
\newblock {\em Photonics and Nanostructures-Fundamentals and Applications}, 52:101071, 2022.

\bibitem{ma2023incorporating}
Yihan Ma, Jonas~Florentin Kolb, Achintha~Avin Ihalage, Andre~Sarker Andy, and Yang Hao.
\newblock Incorporating meta-atom interactions in rapid optimization of large-scale disordered metasurfaces based on deep interactive learning.
\newblock {\em Advanced Photonics Research}, 4(4):2200099, 2023.

\bibitem{wang2019robust}
Evan~W Wang, David Sell, Thaibao Phan, and Jonathan~A Fan.
\newblock Robust design of topology-optimized metasurfaces.
\newblock {\em Optical Materials Express}, 9(2):469--482, 2019.

\bibitem{liu2020big}
Ke~Liu, Larissa~S Novelino, Paolo Gardoni, and Glaucio~H Paulino.
\newblock Big influence of small random imperfections in origami-based metamaterials.
\newblock {\em Proceedings of the Royal Society A}, 476(2241):20200236, 2020.

\bibitem{chen2022gan}
Wei Chen, Doksoo Lee, Oluwaseyi Balogun, and Wei Chen.
\newblock Gan-duf: Hierarchical deep generative models for design under free-form geometric uncertainty.
\newblock {\em Journal of Mechanical Design}, 145(1):011703, 2022.

\bibitem{jang2018wavefront}
Mooseok Jang, Yu~Horie, Atsushi Shibukawa, Joshua Brake, Yan Liu, Seyedeh~Mahsa Kamali, Amir Arbabi, Haowen Ruan, Andrei Faraon, and Changhuei Yang.
\newblock Wavefront shaping with disorder-engineered metasurfaces.
\newblock {\em Nature photonics}, 12(2):84--90, 2018.

\bibitem{xu2022emerging}
Mingfeng Xu, Qiong He, Mingbo Pu, Fei Zhang, Ling Li, Di~Sang, Yinghui Guo, Renyan Zhang, Xiong Li, Xiaoliang Ma, et~al.
\newblock Emerging long-range order from a freeform disordered metasurface.
\newblock {\em Advanced Materials}, 34(12):2108709, 2022.

\bibitem{yu2021engineered}
Sunkyu Yu, Cheng-Wei Qiu, Yidong Chong, Salvatore Torquato, and Namkyoo Park.
\newblock Engineered disorder in photonics.
\newblock {\em Nature Reviews Materials}, 6(3):226--243, 2021.

\bibitem{kao2012}
T.~S. Kao, E.~T.~F. Rogers, J.~Y. Ou, and N.~I. Zheludev.
\newblock “digitally” addressable focusing of light into a subwavelength hot spot.
\newblock {\em Nano Lett.}, 12(6):2728--2731, 2012.

\bibitem{lee2021dynamic}
Doksoo Lee, Shizhou Jiang, Oluwaseyi Balogun, and Wei Chen.
\newblock Dynamic control of plasmonic localization by inverse optimization of spatial phase modulation.
\newblock {\em ACS Photonics}, 9(2):351--359, 2021.

\bibitem{kim2022concurrent}
Myungjoon Kim, Nayoung Kim, and Jonghwa Shin.
\newblock Concurrent inverse design of structured light and metasurface for nanopatterning process.
\newblock In {\em Frontiers in Optics}, pages FM5C--6. Optica Publishing Group, 2022.

\bibitem{you2022learning}
Huaiqian You, Quinn Zhang, Colton~J Ross, Chung-Hao Lee, and Yue Yu.
\newblock Learning deep implicit fourier neural operators (ifnos) with applications to heterogeneous material modeling.
\newblock {\em Computer Methods in Applied Mechanics and Engineering}, 398:115296, 2022.

\bibitem{you2022physics}
Huaiqian You, Quinn Zhang, Colton~J Ross, Chung-Hao Lee, Ming-Chen Hsu, and Yue Yu.
\newblock A physics-guided neural operator learning approach to model biological tissues from digital image correlation measurements.
\newblock {\em Journal of Biomechanical Engineering}, 144(12):121012, 2022.

\bibitem{liu2023ino}
Ning Liu, Yue Yu, Huaiqian You, and Neeraj Tatikola.
\newblock Ino: Invariant neural operators for learning complex physical systems with momentum conservation.
\newblock In {\em International Conference on Artificial Intelligence and Statistics}, pages 6822--6838. PMLR, 2023.

\bibitem{li2020neural}
Zongyi Li, Nikola Kovachki, Kamyar Azizzadenesheli, Burigede Liu, Kaushik Bhattacharya, Andrew Stuart, and Anima Anandkumar.
\newblock Neural operator: Graph kernel network for partial differential equations.
\newblock {\em arXiv preprint arXiv:2003.03485}, 2020.

\bibitem{li2020fourier}
Zongyi Li, Nikola~Borislavov Kovachki, Kamyar Azizzadenesheli, Kaushik Bhattacharya, Andrew Stuart, and Anima Anandkumar.
\newblock Fourier {N}eural {O}perator for {P}arametric {P}artial {D}ifferential {E}quations.
\newblock In {\em International Conference on Learning Representations}, 2020.

\bibitem{lu2019deeponet}
Lu~Lu, Pengzhan Jin, and George~Em Karniadakis.
\newblock Deeponet: Learning nonlinear operators for identifying differential equations based on the universal approximation theorem of operators.
\newblock {\em arXiv preprint arXiv:1910.03193}, 2019.

\bibitem{gupta2021multiwavelet}
Gaurav Gupta, Xiongye Xiao, and Paul Bogdan.
\newblock Multiwavelet-based operator learning for differential equations.
\newblock {\em Advances in neural information processing systems}, 34:24048--24062, 2021.

\bibitem{you2022nonlocal}
Huaiqian You, Yue Yu, Marta D'Elia, Tian Gao, and Stewart Silling.
\newblock Nonlocal kernel network ({NKN}): {A} stable and resolution-independent deep neural network.
\newblock {\em Journal of Computational Physics}, page arXiv preprint arXiv:2201.02217, 2022.

\bibitem{kovachki2021neural}
Nikola Kovachki, Zongyi Li, Burigede Liu, Kamyar Azizzadenesheli, Kaushik Bhattacharya, Andrew Stuart, and Anima Anandkumar.
\newblock Neural operator: Learning maps between function spaces.
\newblock {\em arXiv preprint arXiv:2108.08481}, 2021.

\bibitem{zhang2023metano}
Lu~Zhang, Huaiqian You, Tian Gao, Mo~Yu, Chung-Hao Lee, and Yue Yu.
\newblock Metano: How to transfer your knowledge on learning hidden physics.
\newblock {\em arXiv preprint arXiv:2301.12095}, 2023.

\bibitem{ronneberger2015u}
Olaf Ronneberger, Philipp Fischer, and Thomas Brox.
\newblock U-net: Convolutional networks for biomedical image segmentation.
\newblock In {\em Medical Image Computing and Computer-Assisted Intervention--MICCAI 2015: 18th International Conference, Munich, Germany, October 5-9, 2015, Proceedings, Part III 18}, pages 234--241. Springer, 2015.

\bibitem{liu2020topological}
Zhaocheng Liu, Zhaoming Zhu, and Wenshan Cai.
\newblock Topological encoding method for data-driven photonics inverse design.
\newblock {\em Optics express}, 28(4):4825--4835, 2020.

\bibitem{chan2022remixing}
Yu-Chin Chan, Daicong Da, Liwei Wang, and Wei Chen.
\newblock Remixing functionally graded structures: data-driven topology optimization with multiclass shape blending.
\newblock {\em Structural and Multidisciplinary Optimization}, 65(5), April 2022.

\bibitem{singh2020far}
Anshuman Singh, James~T Hugall, Gaetan Calbris, and Niek~F van Hulst.
\newblock Far-field control of nanoscale hotspots by near-field interference.
\newblock {\em ACS Photonics}, 7(9):2381--2389, 2020.

\bibitem{buijs2021programming}
Robin~D Buijs, Tom~AW Wolterink, Giampiero Gerini, Ewold Verhagen, and A~Femius Koenderink.
\newblock Programming metasurface near-fields for nano-optical sensing.
\newblock {\em Advanced Optical Materials}, 9(15):2100435, 2021.

\bibitem{yoo2023switching}
Hajun Yoo, Hyunwoong Lee, Seongmin Im, Sukhyeon Ka, Gwiyeong Moon, Kyungnam Kang, and Donghyun Kim.
\newblock Switching on versatility: Recent advances in switchable plasmonic nanostructures.
\newblock {\em Small Science}, page 2300048, 2023.

\bibitem{erdman2015functional}
John~M Erdman.
\newblock Functional analysis and operator algebras: An introduction.
\newblock {\em Version October}, 4, 2015.

\bibitem{augenstein2023neural}
Yannick Augenstein, Taavi Repan, and Carsten Rockstuhl.
\newblock Neural operator-based surrogate solver for free-form electromagnetic inverse design.
\newblock {\em ACS Photonics}, 2023.

\bibitem{wen2022u}
Gege Wen, Zongyi Li, Kamyar Azizzadenesheli, Anima Anandkumar, and Sally~M Benson.
\newblock U-fno—an enhanced fourier neural operator-based deep-learning model for multiphase flow.
\newblock {\em Advances in Water Resources}, 163:104180, 2022.

\bibitem{lu2022multifidelity}
Lu~Lu, Rapha{\"e}l Pestourie, Steven~G Johnson, and Giuseppe Romano.
\newblock Multifidelity deep neural operators for efficient learning of partial differential equations with application to fast inverse design of nanoscale heat transport.
\newblock {\em Physical Review Research}, 4(2):023210, 2022.

\bibitem{li2023modeling}
Zhijie Li, Wenhui Peng, Zelong Yuan, and Jianchun Wang.
\newblock Modeling long-term large-scale dynamics of turbulence by implicit u-net enhanced fourier neural operator.
\newblock {\em arXiv preprint arXiv:2305.10215}, 2023.

\bibitem{benitez2023fine}
Jose Antonio~Lara Benitez, Takashi Furuya, Florian Faucher, Xavier Tricoche, and Maarten~V de~Hoop.
\newblock Fine-tuning neural-operator architectures for training and generalization.
\newblock {\em arXiv preprint arXiv:2301.11509}, 2023.

\bibitem{haber2018learning}
Eldad Haber, Lars Ruthotto, Elliot Holtham, and Seong-Hwan Jun.
\newblock Learning across scales---multiscale methods for convolution neural networks.
\newblock In {\em Proceedings of the AAAI Conference on Artificial Intelligence}, volume~32, 2018.

\bibitem{li2022fourier}
Zongyi Li, Daniel~Zhengyu Huang, Burigede Liu, and Anima Anandkumar.
\newblock Fourier neural operator with learned deformations for {PDEs} on general geometries.
\newblock {\em arXiv preprint arXiv:2207.05209}, 2022.

\bibitem{liu2023domain}
Ning Liu, Siavash Jafarzadeh, and Yue Yu.
\newblock Domain agnostic fourier neural operators.
\newblock {\em arXiv preprint arXiv:2305.00478}, 2023.

\bibitem{kingma2014adam}
Diederik~P Kingma and Jimmy Ba.
\newblock Adam: A method for stochastic optimization.
\newblock {\em arXiv preprint arXiv:1412.6980}, 2014.

\bibitem{goodfellow2016deep}
Ian Goodfellow, Yoshua Bengio, and Aaron Courville.
\newblock {\em Deep learning}.
\newblock MIT press, 2016.

\bibitem{buchnev2015electrically}
Oleksandr Buchnev, Nina Podoliak, Malgosia Kaczmarek, Nikolay~I Zheludev, and Vassili~A Fedotov.
\newblock Electrically controlled nanostructured metasurface loaded with liquid crystal: toward multifunctional photonic switch.
\newblock {\em Advanced Optical Materials}, 3(5):674--679, 2015.

\bibitem{li20214d}
Bing Li, Chao Zhang, Fang Peng, Wenzhi Wang, Bryan~D Vogt, and KT~Tan.
\newblock 4d printed shape memory metamaterial for vibration bandgap switching and active elastic-wave guiding.
\newblock {\em Journal of Materials Chemistry C}, 9(4):1164--1173, 2021.

\bibitem{wang2023physics}
Liwei Wang, Yilong Chang, Shuai Wu, Ruike~Renee Zhao, and Wei Chen.
\newblock Physics-aware differentiable design of magnetically actuated kirigami for shape morphing.
\newblock {\em arXiv preprint arXiv:2308.05054}, 2023.

\bibitem{wang2023inverse}
Chao Wang, Zhi Zhao, and Xiaojia~Shelly Zhang.
\newblock Inverse design of magneto-active metasurfaces and robots: Theory, computation, and experimental validation.
\newblock {\em Computer Methods in Applied Mechanics and Engineering}, 413:116065, 2023.

\bibitem{malek2017strain}
Stephanie~C Malek, Ho-Seok Ee, and Ritesh Agarwal.
\newblock Strain multiplexed metasurface holograms on a stretchable substrate.
\newblock {\em Nano letters}, 17(6):3641--3645, 2017.

\bibitem{bilal2017reprogrammable}
Osama~R Bilal, Andr{\'e} Foehr, and Chiara Daraio.
\newblock Reprogrammable phononic metasurfaces.
\newblock {\em Advanced materials}, 29(39):1700628, 2017.

\bibitem{xu2019stretchable}
Zefeng Xu and Yu-Sheng Lin.
\newblock A stretchable terahertz parabolic-shaped metamaterial.
\newblock {\em Advanced Optical Materials}, 7(19):1900379, 2019.

\bibitem{marler2010weighted}
R~Timothy Marler and Jasbir~S Arora.
\newblock The weighted sum method for multi-objective optimization: new insights.
\newblock {\em Structural and multidisciplinary optimization}, 41:853--862, 2010.

\bibitem{sener2018multi}
Ozan Sener and Vladlen Koltun.
\newblock Multi-task learning as multi-objective optimization.
\newblock {\em Advances in neural information processing systems}, 31, 2018.

\bibitem{paszke2017automatic}
Adam Paszke, Sam Gross, Soumith Chintala, Gregory Chanan, Edward Yang, Zachary DeVito, Zeming Lin, Alban Desmaison, Luca Antiga, and Adam Lerer.
\newblock Automatic differentiation in pytorch.
\newblock 2017.

\bibitem{lee2023t}
Doksoo Lee, Yu-Chin Chan, Wei Chen, Liwei Wang, Anton van Beek, and Wei Chen.
\newblock t-metaset: Task-aware acquisition of metamaterial datasets through diversity-based active learning.
\newblock {\em Journal of Mechanical Design}, 145(3):031704, 2023.

\bibitem{Chan2022Yu-ChinDissertation}
Yu-Chin Chan.
\newblock {Yu-Chin Chan PhD Dissertation}.
\newblock Technical report, 2022.

\bibitem{mcinnes2018umap}
Leland McInnes, John Healy, and James Melville.
\newblock Umap: Uniform manifold approximation and projection for dimension reduction.
\newblock {\em arXiv preprint arXiv:1802.03426}, 2018.

\bibitem{shaltout2019spatiotemporal}
Amr~M Shaltout, Vladimir~M Shalaev, and Mark~L Brongersma.
\newblock Spatiotemporal light control with active metasurfaces.
\newblock {\em Science}, 364(6441):eaat3100, 2019.

\bibitem{kang2019recent}
Lei Kang, Ronald~P Jenkins, and Douglas~H Werner.
\newblock Recent progress in active optical metasurfaces.
\newblock {\em Advanced Optical Materials}, 7(14):1801813, 2019.

\bibitem{balogun2019optically}
Oluwaseyi Balogun.
\newblock Optically detecting acoustic oscillations at the nanoscale: Exploring techniques suitable for studying elastic wave propagation.
\newblock {\em IEEE Nanotechnology Magazine}, 13(3):39--54, 2019.

\bibitem{comsol2020}
{COMSOL AB}.
\newblock {\em {COMSOL Multiphysics® v.5.6}}.
\newblock Stockholm, Sweden, 2020.
\newblock Available from: \url{https://www.comsol.com}.

\bibitem{jin2003efficient}
Ruichen Jin, Wei Chen, and Agus Sudjianto.
\newblock An efficient algorithm for constructing optimal design of computer experiments.
\newblock In {\em International design engineering technical conferences and computers and information in engineering conference}, volume 37009, pages 545--554, 2003.

\bibitem{ren2021survey}
Pengzhen Ren, Yun Xiao, Xiaojun Chang, Po-Yao Huang, Zhihui Li, Brij~B Gupta, Xiaojiang Chen, and Xin Wang.
\newblock A survey of deep active learning.
\newblock {\em ACM Computing Surveys (CSUR)}, 54(9):1--40, 2021.

\bibitem{chen2020design}
M.~Chen, J.~Jiang, and J.~A. Fan.
\newblock Design space reparameterization enforces hard geometric constraints in inverse-designed nanophotonic devices.
\newblock {\em ACS Photonics}, 7(8):2039--2046, 2020.

\bibitem{hammond2021photonic}
Alec~M. Hammond et~al.
\newblock Photonic topology optimization with semiconductor-foundry design-rule constraints.
\newblock {\em Optics Express}, 29(15):23916--23938, 2021.

\bibitem{tanriover2022deep}
Ibrahim Tanriover, Doksoo Lee, Wei Chen, and Koray Aydin.
\newblock Deep generative modeling and inverse design of manufacturable free-form dielectric metasurfaces.
\newblock {\em ACS Photonics}, 2022.

\bibitem{wang2021data}
Liwei Wang, Siyu Tao, Ping Zhu, and Wei Chen.
\newblock Data-driven topology optimization with multiclass microstructures using latent variable gaussian process.
\newblock {\em Journal of Mechanical Design}, 143(3):031708, 2021.

\bibitem{zhu2023reliable}
Min Zhu, Handi Zhang, Anran Jiao, George~Em Karniadakis, and Lu~Lu.
\newblock Reliable extrapolation of deep neural operators informed by physics or sparse observations.
\newblock {\em Computer Methods in Applied Mechanics and Engineering}, 412:116064, 2023.

\bibitem{snoek2012}
J.~Snoek, H.~Larochelle, and R.~P. Adams.
\newblock Practical bayesian optimization of machine learning algorithms.
\newblock In {\em Adv. Neural Inf. Process. Syst.}, volume~4, pages 2951--2959, 2012.

\bibitem{swersky2013multi-task}
Kevin Swersky, Jasper Snoek, and Ryan~P. Adams.
\newblock Multi-task bayesian optimization.
\newblock In {\em Advances in Neural Information Processing Systems 26}, 2013.

\bibitem{halim2021performance}
A.~Hanif Halim, Idris Ismail, and Swagatam Das.
\newblock Performance assessment of the metaheuristic optimization algorithms: an exhaustive review.
\newblock {\em Artificial Intelligence Review}, 54:2323--2409, 2021.

\bibitem{shahriari2015taking}
Bobak Shahriari, Kevin Swersky, Ziyu Wang, Ryan~P Adams, and Nando De~Freitas.
\newblock Taking the human out of the loop: A review of bayesian optimization.
\newblock {\em Proceedings of the IEEE}, 104(1):148--175, 2015.

\bibitem{snoek2012practical}
Jasper Snoek, Hugo Larochelle, and Ryan~P Adams.
\newblock Practical bayesian optimization of machine learning algorithms.
\newblock {\em Advances in neural information processing systems}, 25, 2012.

\bibitem{comsol2021}
Comsol multiphysics v. 5.3.
\newblock \url{https://www.comsol.com}, 2021.
\newblock Accessed: 2021-12-12.

\end{thebibliography}
\bibliographystyle{unsrt}


\end{document}